\documentclass[preprint,12pt]{elsarticle}
\usepackage[letterpaper,top=2cm,bottom=2cm,left=3cm,right=3cm,marginparwidth=1.75cm]{geometry}
\usepackage{amssymb}
\usepackage{amsmath}
\usepackage{graphicx}
\usepackage{dirtytalk}
\usepackage[colorlinks=true,linkcolor=blue]{hyperref}
\usepackage{multicol}
\usepackage{natbib}
\usepackage{comment}
\usepackage{caption}
\usepackage{subcaption}
\usepackage{algorithm}
\usepackage{algpseudocode}
\usepackage{physics}
\usepackage[autostyle]{csquotes}
\usepackage{dirtytalk}
\usepackage{lineno}

\newcommand{\mbs}[1]{\boldsymbol{#1}}

\def\bA{{\mbs{A}}}  \def\bC{{\mbs{C}}}
  \def\bF{{\mbs{F}}}
\def\bG{{\mbs{G}}}

\def\bP{{\mbs{P}}}  
  
 \def\bX{{\mbs{X}}} \def\bY{{\mbs{Y}}}
 \def\b0{{\mbs{0}}}
\def\ba{{\mbs{a}}}

 \def\bn{{\mbs{n}}} 
  
  \def\bu{{\mbs{u}}}
 \def\bx{{\mbs{x}}} \def\by{{\mbs{y}}}

\def \bsig	{\mbs{\sigma}}

\def    \tr {\textrm{tr\,}}

\begin{document}

\title{A self-contact electromechanical framework for intestinal motility}

\author[inst1]{René Thierry Djoumessi\corref{cor1}}
\author[inst1]{Pietro Lenarda\corref{cor1}}
\author[inst2]{Alessio Gizzi}
\author[inst1]{Marco Paggi}

\affiliation[inst1]{organization={IMT School for Advanced Studies Lucca},
            addressline={Piazza San Francesco 19}, 
            city={Lucca},
            postcode={55100}, 
            country={Italy}}

\affiliation[inst2]{organization={Department of Engineering, Università Campus Bio-Medico di Roma},
            addressline={Via A. del Portillo 21}, 
            city={Rome},
            postcode={00128}, 
            country={Italy}}

\cortext[cor1]{Corresponding author Email addresses: rene.djoumessi@imtlucca.it (R. T. Djoumessi, pietro.lenarda@imtlucca.it (P. Lenarda))}

\begin{frontmatter}

\begin{abstract}
This study introduces an advanced multiphysics and multiscale modeling approach to investigate intestinal motility. We propose a generalized electromechanical framework that incorporates contact mechanics, enabling the development of a unique and innovative model for intestinal motility. The theoretical framework includes an electromechanical model coupling a microstructural material model, which describes the intestinal structure, with an electrophysiological model that captures the propagation of slow waves. Additionally, it integrates a self-contact detection algorithm based on a nearest-neighbour search and the penalty method, along with boundary conditions that account for the influence of surrounding organs. A staggered finite element scheme implemented in FEniCS is employed to solve the governing equations using the finite element method. The model is applied to study cases of moderate and severe strangulation hernia, as well as intestinal adhesion syndrome. The results demonstrate that low peristalsis takes place in the pre-strangulation zone. At the same time, very high pressure is recorded in the strangulation zone, and peristaltic contractions persisted in the healthy region. For adhesions, the results indicate a complete absence of peristalsis in the adherent region. The model successfully reproduces both qualitatively and quantitatively propagative contractions in complex scenarios, such as pre- and post-surgical conditions, thereby highlighting its potential to provide valuable insights for clinical applications.

\end{abstract}


\end{frontmatter}


\section{Introduction}
The intestine, a key part of the digestive tract, is crucial in processing food and absorbing nutrients. It is divided into two main sections: the small intestine and the large intestine. The small intestine is responsible for the majority of nutrient absorption, breaking down food into essential components such as proteins, carbohydrates, and fats. Once most nutrients have been absorbed, the remaining undigested food residues, or chyme, pass into the large intestine. The large intestine primary function is to absorb water, electrolytes, and vitamins produced by gut bacteria while also compacting waste into the stool for elimination~\citep{azzouz2018physiology, sulaiman2019mri, precup2019gut}.

However, our understanding of the mechanisms behind the different types of movement, such as peristalsis and segmentation, remains incomplete, making its modeling and prediction extremely complex. Over the last two decades, multiphysics and multiscale models have been proposed to better understand intestine electrophysiology \citep{aliev2000simple, sharon2018expertise, du2018progress}, passive mechanics \citep{nagaraja2021phase, sokolis2013microstructure} and active electromechanics \citep{brandstaeter2018computational, du2013model, DJOUMESSI2024116989}. Current effort aims at filling the gap with advanced experimental settings \citep{patton2024simultaneous, athavale2024mapping, kuruppu2024electromechanical}.

Unlike other organs, such as the heart and lungs, the intestine is not protected by the skeleton or the rib cage. This means that the intestine is almost free to move in any direction. However, specific structures in the body exert forces to keep it in place. These include the mesentery and the abdomen, which exert pressure to prevent the intestine from ‘falling’ into the abdominal cavity \citep{byrnes2019anatomy, sensoy2021review}. To the best of our knowledge, there are not yet mechanical or electromechanical models that consider these interactions when modeling intestine motility--as done for the heart \citep{baier2018modelling, propp2020orthotropic}. In addition, recurrent post-surgical pathologies (more than 9 out of 10 patients who had abdominal surgery \citep{tabibian2017abdominal}) are characterized by bands of scar fibrous tissues or by local adhesions that can lead to increased stiffness and reduced motility \citep{frager1996detection, cardoz2024acute, furtado2024estrangulated}. Hernia is also one of the most serious pathologies in which part of the intestine penetrates another abdominal cavity, causing intestinal strangulation \citep{jadhav2022prospective, moazzez2021outcomes, klifto2021risk}.

Whether the pressure exerted by other organs on the intestine or the pathologies mentioned, these factors have a particular effect: they can lead to a form of self-contact or adhesions between the different tracts of the intestine \citep{akbaba2021multilayer}. 
Several algorithms have been developed in the literature to study contact between solids, both in small and large deformations, as well as self-contact \citep{chouly2023contact,li2022contact,areias2023continuous}. Most of these methods are based on the concept of ‘master’ and ‘slave’ boundaries to describe the interactions between the contact surfaces \citep{xuan2024penalty}. Among the approaches used to impose non-penetration between these surfaces, one of the most widespread and most straightforward to implement is the penalty method \citep{pore2021review, bozorgmehri2021study}. This method introduces an artificial constraint proportional to the penetration depth, thus guaranteeing compliance with the non-interpenetration condition. The penalty approach is easy to implement in finite element codes while offering robustness for contact problems \citep{bozorgmehri2021study}. However, its effectiveness may depend on the appropriate choice of the penalty parameter, which must be high enough to avoid significant interpenetration without introducing excessive rigidity into the system.

This study proposes a generalized electromechanical framework of the intestinal system incorporating self-contact. The suggested model will also allow for an analysis \emph{in-silico} of specific pathologies, providing a starting point for a better understanding of the effects of a pathology or surgical intervention on the overall motility of the intestinal tract.

The manuscript is organized as follows. 
In Section \ref{sec:model}, the generalized electromechanical framework for intestine motility is recalled based on our previous work \citep{DJOUMESSI2024116989} considering a self-contact formulation. 
In Section \ref{sec:Numerical}, the strong form of the problem is derived, and the discretization of finite elements is presented.
In Section \ref{sec:numerical application}, numerical experiments are carried out, as well as the use of the model to investigate the effect of surrounding organs on GI motility in two typical conditions, namely herniation and adherence syndrome.
Conclusions, limitations, and perspectives are discussed in Section \ref{sec:conclusion}.

\section{Colon electromechanics and self-contact}
\label{sec:model}
In this section, we briefly recall the governing equations for active strain finite deformations coupled with GI electrophysiology.

We represent a scalar, a vector, and a second-order tensor with the lowercase letters ($a$), lowercase bold letters ($\ba$), and capital bold letters ($\bA$), respectively, and ($\bA^T$) stands for the transpose of a tensor. According to the tensor notation, we indicate the scalar product with $(\cdot)$, the double contraction with $(:)$, and the dyadic product with $(\otimes)$. Moreover, $\nabla$, $\nabla\cdot$ and $\nabla^2$ represent the gradient, divergence, and Laplace operator, respectively.

\subsection{Finite kinematics}
Kinematics of deformable GI tissue is embedded in the classical description of continuum mechanics under the assumption of finite elasticity \citep{holzapfel2002nonlinear}. 

Let $\bX, \ \bx$ be the material position vector in the undeformed and deformed configuration $\Omega_0, \ \Omega_t\subset R^d$, $d=2, 3$ respectively, the deformation gradient tensor and its associated Jacobian are denoted as $\bF={\partial\bx}/{\partial\bX}$ and $J=\det\bF > 0$, the left Cauchy-Green deformation tensor with $\bC=\bF^T \bF$, the first isotropic invariant of deformation with $I_1(\bC)=\tr(\bC)$, where $\tr( \cdot )$ is the trace operator, and the fourth anisotropic pseudo-invariant is $I_4(\bC)=\bC:\bG$, where $\bG$ denotes the structure tensor.

The contraction of the intestine combines active and passive behaviours, coupling electrophysiological cellular dynamics with a hyperelastic response of the material in a nonlinear manner. The active strain approach \citep{brandstaeter2018computational, cherubini2008electromechanical, ambrosi2011electromechanical,ruiz2020thermo} remains an effective way for this type of coupling. In particular, the deformation gradient tensor is multiplicatively decomposed into an elastic part, $\bF_e$, and an inelastic part, $\bF_a$ as $ \bF = \bF_e \bF_a \,$. The reader can refer to our previous article \citep{DJOUMESSI2024116989} for details on $\bF_a$ and $\bF_e$.

\subsection{Electrophysiology}
We consider a simplified phenomenological model of intestine electrophysiology. The smooth muscle cells (SMC) and interstitial cells of Cajal (ICC) layers are denoted by the indices $s$ and $i$, respectively. The resulting nonlinear partial differential equations for the coupled reaction-diffusion system describing the interaction between transmembrane potential variables, $u_s, u_i$, and slow current variables, $v_s, v_i$ is given by:
\begin{linenomath}
\begin{subequations}
\begin{align}
\frac{\partial u_s}{\partial t} &= f(u_s)+D_s\nabla^2 u_s-v_s + F_s(u_s,u_i) \quad \textrm{on} \quad \Omega_0 \times [0,T], \label{eq:14a}
\\
\frac{\partial v_s}{\partial t}  &= \epsilon_s[\lambda_s (u_s-\beta_s)-v_s] \quad \textrm{on} \quad \Omega_0 \times [0,T], \\
\frac{\partial u_i}{\partial t} &= g(u_i)+D_i\nabla^2 u_i-v_i + F_i(u_s,u_i) + I_{stim}^i \quad \textrm{on} \quad \Omega_0 \times [0,T], \label{eq:14c}\\
\frac{\partial v_i}{\partial t}  &= \epsilon_i(z)[\lambda_i (u_i-\beta_i)-v_i] \quad \textrm{on} \quad \Omega_0 \times [0,T],
\end{align}
\label{electrophysiology}
\end{subequations}
\end{linenomath}
where:
\begin{linenomath}
\begin{subequations}
\begin{align}
f(u_s) &=  k_su_s(u_s-a_s)(1-u_s) \,, \quad
& F_s(u_s,u_i) = \alpha_s D_{si}(u_s-u_i) \,, \\
g(u_i) &=  k_iu_i(u_i-a_i)(1-u_i) \,, \quad
& F_i(u_s,u_i) = \alpha_i D_{is}(u_s-u_i) \,.
\end{align}
\label{electrophysiologyfun}
\end{subequations}
\end{linenomath}
Here, $I_{stim}^i$ is the external currents applied to the ICC; $D_s, D_i$ are the diffusivities; $\lambda_s, \lambda_i$ are the coupling factors between the membrane potential and recovery variable; $D_{si}, D_{is}$ are the diffusivities of the gap junctions between the two cell species; $k_i, k_s, a_s, a_i,\alpha_s, \alpha_i$ are phenomenological model parameters and their values are provided in \ref{sec:A}.
The parameter $\epsilon(z)$ represents a space-dependent excitability function, decreasing with distance from the pylorus $z$~\citep{aliev2000simple}. More information concerning the model parameters can be found in \citep{aliev2000simple, gizzi2010electrical, DJOUMESSI2024116989}.

\subsection{Constitutive mechanical model}
\begin{figure}[h]
\centering\includegraphics[width=\textwidth]{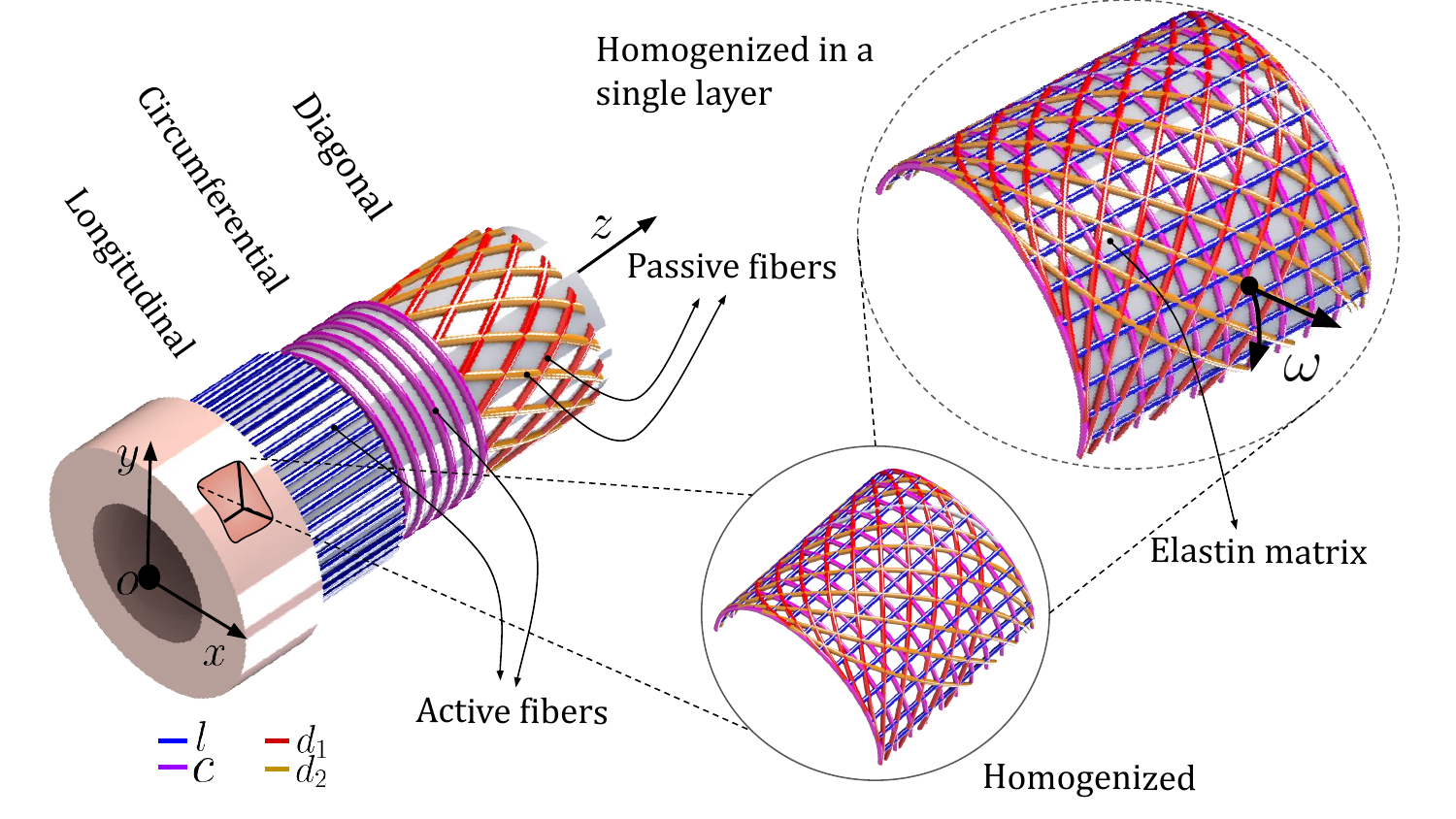}
    \caption{Schematic segment of the intestine. The zoomed cross-section represents the homogenized fiber microstructure, which is composed of four families of fibers embedded in an isotropic elastin matrix. The directions of the fibers are uniquely defined with respect to the circumferential direction by the angle $\omega$; $l$ represents the external longitudinal muscular layer, $c$ the internal circumferential muscular fiber, $d_1$ and $d_2$ are the submucosa diagonally collagen fibers.}
    \label{fig:1}
\end{figure}

Following usual formulations for soft tissue biomechanics, we consider the intestinal wall as an anisotropic incompressible material. The tissue is reinforced by four families of fibers with specific preferential directions (see Fig.~\ref{fig:1} for details). Accordingly, the strain energy density $\Psi$ comprises isotropic, $\Psi^{\rm iso}$, and anisotropic, $\Psi^{\rm aniso}$, contributions. For the sake of simplicity, the isotropic part is considered of neo-Hookean type, while the anisotropic part is modeled as an Ogden-Holzapfel structure-based energy density accounting both for passive and active components. The passive part is associated with the mechanical response of directional collagen fibers in the submucosal layer $(d_1, d_2)$, while the active contribution is due to the presence of SMC fibers in the longitudinal $(l)$, and circumferential $(c)$, directions and is the only part that contributes in the definition of the active part of the deformation gradient tensor $\bF_a$ (see \citep{DJOUMESSI2024116989}):
\begin{linenomath}
\begin{equation} 
\label{eq:consti}
  \Psi =
  \Psi^{\rm iso} +
  \Psi^{\rm aniso}  
  = 
  \mu(I_1 -3)+
  \sum_{i \in \{ l, c, d_1, d_2 \}} \frac{k_1^i}{4k_2^i}[e^
  {k_2^i(I_4^i -1)_+^2}-1] - p(J-1)  
  .
\end{equation}
\end{linenomath}
Here, the notation $(y)_+:= y$ if $y\geq 0$ reproduces the tension-compression switch approximation \citep{patel2022biomechanical} and the anisotropic fourth invariant $I_4^j = \bC:(\bn_j \otimes \bn_j)$ is distinguished for each fiber family $j \in \{l,c,d_1,d_2 \}$. The material parameters  $k_1^j$ (stiffness-like),  $k_2^j$ (nondimensional) are associated with the directional behavior of the material, $\mu$ is the passive isotropic stiffness and $p$ stands for solid hydrostatic pressure.

\subsection{Modeling of the self-contact and active strain approach}
\subsubsection{Penalty approach to contact }
Contact problems pose theoretical and numerical challenges, especially in large deformation, where complex geometric and mechanical quantities depend on an a priori unknown correspondence between contact surfaces \citep{mlika2017unbiased}. To address these problems, several methods have been developed, the most commonly used being the node-to-surface (NTS) approach in a master-slave configuration \citep{laursen1993continuum, paggi2011contact, wriggers2006computational}. However, this configuration presents major difficulties in the cases of self-contact and multi-body contact, where it is impossible or impractical to designate a master and a slave surface a priori.

To avoid these difficulties, some unbiased formulations for contact have been proposed \citep{sauer2015unbiased, neto2020numerical}. In this work, an unbiased (no master and slave concept) version based on a penalty approach in large deformation has been adopted.

\begin{figure}[h]
    \centering
    \includegraphics[width=\textwidth]{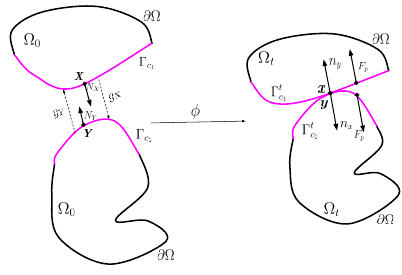}
    \caption{Concept of contact. $\bX$ and $\bY$ are material points in the reference configuration, while $\bx$ and $\by$ represent their positions in the current configuration. $\Gamma_{c_i}$ and $\Gamma_{c_i}^t$ are the contact surfaces in the reference and current configurations, respectively with $i \in {1, 2}$. $F_p$ denotes the penalty force applied to both contact surfaces and $\phi$ is the deformation map.}
    \label{contactconfk}
\end{figure}

Our strategy (see Fig.~\ref{contactconfk}) aims to prevent any interpenetration between the two surfaces $\Gamma_c$ in contact during simulation. To this purpose, we proposed a method based on the calculation of the Euclidean distance between the contact surfaces via a contact search method based on k-d-nearest neighbors \citep{hou2018advanced, kamaludin2023efficient}. When a distance below a certain threshold is detected, the penalty force is applied to prevent the interpenetration of the surfaces. This method introduces a corrective force proportional to the contact violation. This approach effectively models contact interactions by maintaining a strict separation between the surfaces while ensuring robust convergence of the numerical solutions.

The aim of this approach is to maintain a positive gap, by penalizing any violation of the interpenetration condition, thus ensuring precise contact management by avoiding any physical overlapping of surfaces.

Mathematically, the contact conditions on the two contact surfaces are given by:
\begin{linenomath}
    \begin{subequations}
    \begin{align}
        g_X(N_X) = \bu_X \cdot N_X \geq 0;\quad \bP_{N_X} \geq 0;\quad \bP_{N_X}\cdot g_X(N_X) = 0;\quad {\rm on} \quad \Gamma_{c_1} \,,\\
        g_Y(N_Y) = \bu_Y \cdot N_Y \geq 0;\quad \bP_{N_Y} \geq 0;\quad \bP_{N_Y}\cdot g_Y(N_Y) = 0;\quad {\rm on} \quad \Gamma_{c_2} \,.
        \end{align}
        \label{signorini}
    \end{subequations}
\end{linenomath}
In Eqs.~\ref{signorini}, $g_X(N_X)$, and $g_Y(N_Y)$ are the gaps between two generic points entering in contact, $\bP_{N_X}$ and $\bP_{N_Y}$ are the normal first Piola-Kirchhoff stress tensor aligned with the normals $N_X$ and $N_Y$ respectively. One of the  simplest ways to automatically ensure the contact conditions in Eqs.~\ref{signorini} consists of imposing the following expressions for the normal Piola-Kirchhoff stress tensors:
\begin{equation}
\bP_{N_X} = K_0\langle g_X(N_X)\rangle; \quad \bP_{N_Y} = K_0\langle g_Y(N_Y)\rangle \,\; ,
\label{penalty}
\end{equation}
where $\langle \mathcal{J} \rangle = (\mid \mathcal{J}\mid+ \mathcal{J})/2$ is the Mackauley operator computing the positive part of the gaps and $K_0$ is a large stiffness penalty value used to penalize the gaps.

\subsubsection{Contact search algorithm based on k-d tree nearest neighbors}

Search algorithms usually employed for contact mechanics can be computationally expensive. The standard "brute-force" approach for finding the relative distances between two discretized surfaces has, in general, computational complexity $O(N^2)$, as it evaluates the distance between each point on one contact surface and all the points on the other contact surface. To accelerate this algorithm, several strategies have been proposed based on the nature of the contact problem and the description of the contact surfaces \citep{paggi2011contact, paggi2020computational, zavarise1998method, wriggers2025third, chouly2018unbiased}. Recently, a contact algorithm based on the representation of the gap as the solution of the screened Poisson equation has been proposed in \citep{areias2023continuous} and another one representing the gap as a phased field of an Eulerian diffused problem has also been proposed in \citep{lorez2024eulerian}. We introduced a method based on the Nearest-Neighbor k-d algorithm in the present work \citep{hou2018advanced, kamaludin2023efficient}.

In computer science, a k-d trees (short for k-dimensional tree) is a space-partitioning data structure for organizing points in a k-dimensional space concerning exactly k orthogonal axes or a space of any number of dimensions. k-d trees are useful data structures for several applications, such as searches involving a multidimensional search key (e.~g., range searches and nearest neighbor searches) and creating point clouds used in clustering problems. In computer vision, the k-d tree algorithm has been widely used to match key points between images by efficiently finding the nearest neighbors of feature descriptors like SIFT or SURF, enabling tasks such as image-stitching and 3D reconstruction \citep{lowe2004distinctive}. In machine learning, k-d tree supports algorithms like k-nearest neighbors (k-NN) for classification and regression by organizing multidimensional feature spaces for rapid neighbor searches \citep{hou2018advanced, ram2019revisiting}.

In the framework of search of minimal distances between contact surfaces, the k-d optimization restricts the distance calculation to the closest points belonging to the opposite surface, significantly reducing the computational complexity to $O(N \log(N))$, while maintaining sufficient accuracy for modeling contact interactions. The two contact surfaces, called $\Gamma_{c_1}$ and $\Gamma_{c_2}$, belong to a deformable domain $\Omega_{0}$. In the reference configuration, the nodal coordinates of the points on each surface are extracted from the initial mesh. When the structure is subjected to loading, the surfaces move according to the nodal displacements resulting from the resolution of the mechanical problem. To follow the evolution of the gap, we need to update the positions of the points on these surfaces in the deformed configuration.

\begin{figure}[h]
    \centering
    \includegraphics[width=0.9\textwidth]{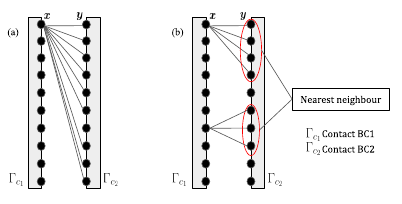}
    \caption{Concept of contact search based on the nearest neighbor algorithm: $\bx$ and $\by$ represent points on the contact surfaces $\Gamma_{c_1}$ and $\Gamma_{c_2}$, respectively, in the current configuration. Panel (a) illustrates the search method without using the nearest neighbor algorithm, while panel (b) highlights the optimized search method incorporating the nearest neighbor algorithm.}
    \label{Euclideancontact}
\end{figure}

An illustrative schematic of the method is given in Fig.~\ref{Euclideancontact}, and described as follows.
To find the contact nodes of $y$ closest to $x$, a hierarchical structure divides the space of $y$ into subregions based on spatial dimensions. This organization forms a binary tree, where each node represents a partobtained the contact nodes of $y$. During the search, we descend the tree by following the subtrees that contain the points most likely to be in contact with $x$. Once a small group of nodes is identified, the contact distance $d(x, y)$ between $x$ and these contact nodes is calculated. During the ascent of the tree, we check if any ignored subtrees could contain closer contact nodes based on the distance to the separating hyperplane. If not, these branches are discarded.
The the minimum distance which represents the local gap between the two contact surfaces is mathematically given by:
\begin{equation}
    g_X(x) = \min_{y \in \Gamma_{c_2}} d(x, y)
    \label{gap}
\end{equation}

The contact search algorithm base on k-d tree nearest neighbour is described in {Algorithm \ref{alg:kd_tree}}.
\begin{algorithm}[H]
\caption{Nearest Neighbor Search Using KD-Tree}
\label{alg:kd_tree}
\begin{algorithmic}[1]
\Require A set of points $\Gamma_{c_2} = \{Y_1, Y_2, \ldots, Y_M\}$, a set of query points $\Gamma_{c_1} = \{X_1, X_2, \ldots, X_N\}$.
\Ensure The minimum distance $g_X(X)$ for each $X \in \Gamma_{c_1}$.

\State \textbf{Step 1: Build the k-d tree}
\State Construct a k-d tree for the points in $\Gamma_{c_2}$.
\Comment{This organizes the points of $\Gamma_{c_2}$ into a space-partitioning tree.}

\State \textbf{Step 2: Initialize Results}
\State Create an empty array $G$ to store the gap $g_X(X)$ for each $X \in \Gamma_{c_1}$.

\State \textbf{Step 3: Perform Nearest Neighbor Search}
\For{each point $X \in \Gamma_{c_1}$}
    \State Query the k-d tree to find the nearest neighbor $Y \in \Gamma_{c_2}$ to $X$.
    \State Compute the Euclidean distance: $g_X(X)$
    
    \State Store $g_X(X)$ in $G$.
\EndFor

\State \textbf{Step 4: Return Results}
\State \Return $G$, the array of minimum distances.

\end{algorithmic}
\end{algorithm}

\section{Numerical implementation and benchmark}
\label{sec:Numerical}

\subsection{Benchmark Test: Hertzian contact problem}
\begin{figure}[h!]
    \centering
    \includegraphics[width=\textwidth]{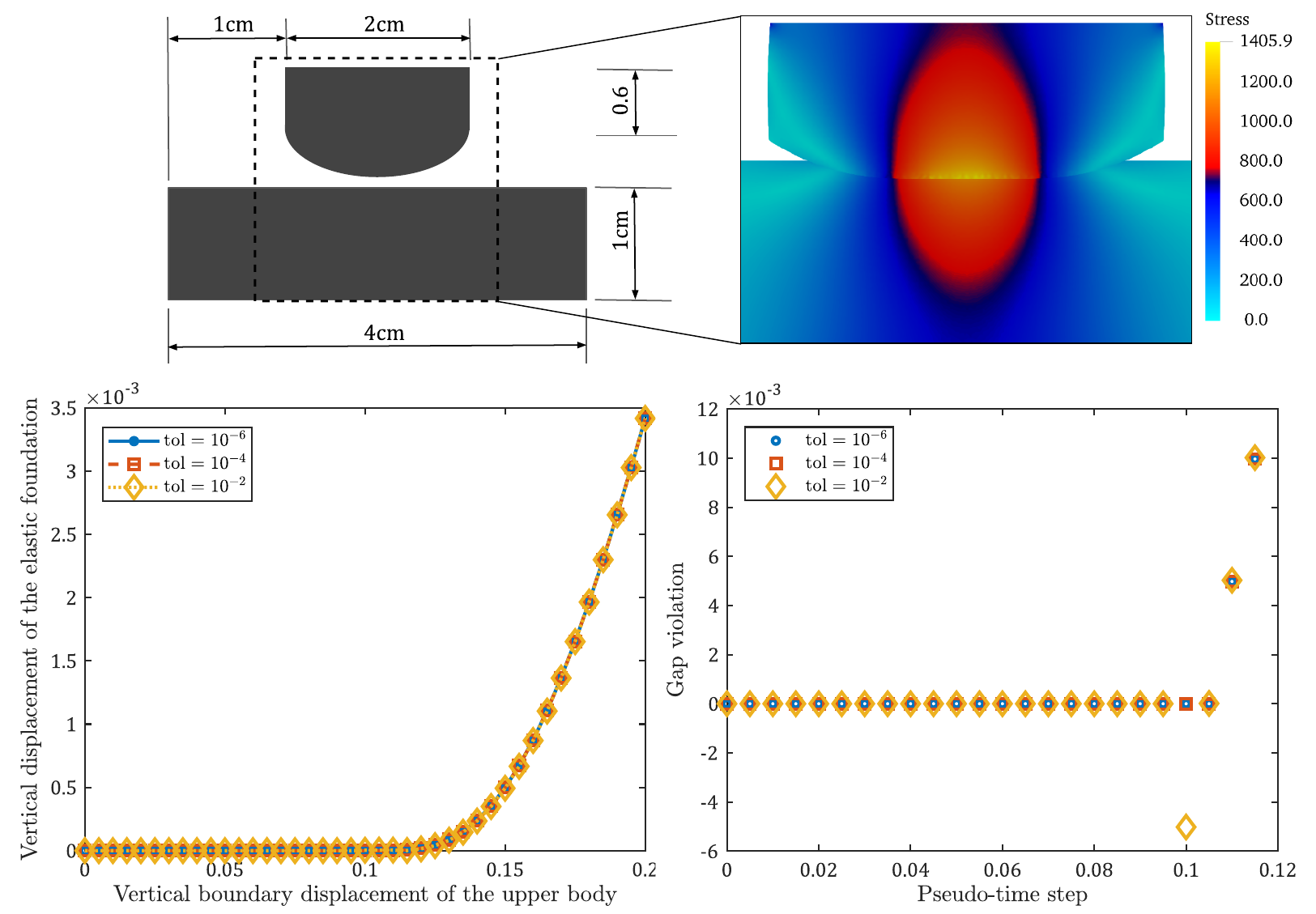}
    \caption{Benchmark problem of the compression of an elastic body against an elastic foundation. The mesh was refined with a local dimension of  0.03 cm for the two bodies. The zoom plot shows the stress, gap violation and vertical displacement of the plate for the benchmark problem. The two solids are modeled as linear elastic bodies with different material properties: $E_u = 7000$ MPa , $\nu_u = 0.3$, $E_f = 70000$ MPa , $\nu_f = 0.45$.}
    \label{benchmarkresult1}
\end{figure}

The contact method is tested against a benchmark problem from Wriggers et al. \citep{wriggers2013finite}. The problem configuration is explained in Fig~.\ref{benchmarkresult1}. In this benchmark, the upper is moved downwards in progressive steps to compress the elastic foundation. We model the upper body and the foundation as linear elastic materials with different material properties. The constitutive law for each solid is characterized by $\bsig = \lambda \textrm{tr}(\epsilon)I + 2\mu \epsilon$, where $\lambda$ and $\mu$ are the Lamé constants and $\epsilon$ is the strain.
The following material properties were $E_u = 7000  \;\rm MPa $, $E_f = 70000  \;\rm MPa $, $\nu_u = 0.3$, and $\nu_f = 0.45$. where the subscripts `u' and `f' stand for the upper body and foundation, respectively.
The stress distribution obtained with the contact method is reported in  Fig.~\ref{benchmarkresult1}. 
We observe that the method exhibits a violation when the compression is maintained beyond a certain threshold. However, this gap violation appears to be minimally affected by the defined tolerance. 
The vertical displacement of the foundation against the vertical displacement of the upper body is shown in Fig.~\ref{benchmarkresult1} as reported in \citep{wriggers2013finite}. The results show little variation with respect to the tolerance, demonstrating the robustness of the method.

\begin{figure}[h!]
    \centering
    \includegraphics[width=\textwidth]{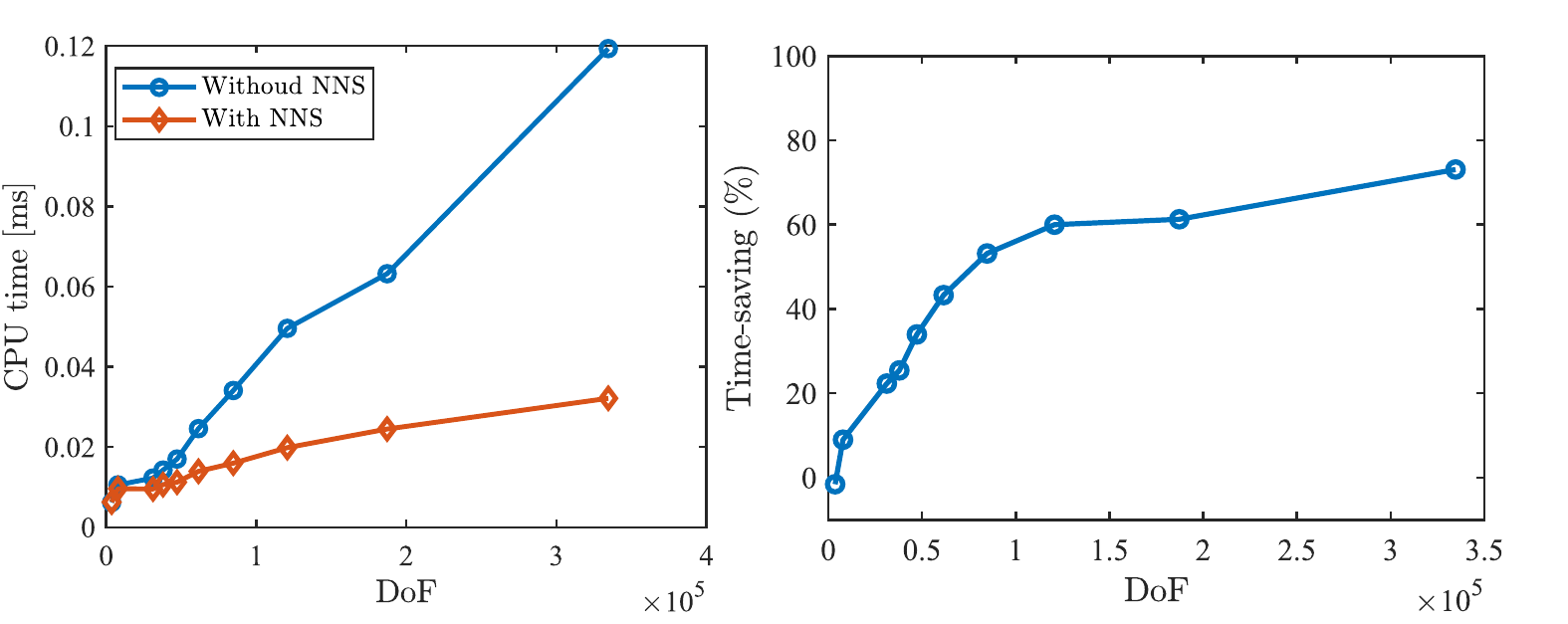}
    \caption{Left: computational time with and without the nearest neighbor contact search (NNS) vs. the number of DoFs. Right: time-saving percentage per degree of freedom.}
    \label{Timec}
\end{figure}

An additional analysis was conducted to evaluate the computational cost of our algorithm. For this purpose, we analyzed the computational time required for calculating the gap in two distinct scenarios: when the nearest neighbor search (NNS) was used and when it was not. The analysis was conducted as a function of the degrees of freedom (DoFs) in the problem. The results show that using the nearest neighbor search significantly reduces the computational time, as illustrated in Fig.~\ref{Timec}, which also presents the time-saving percentage per degree of freedom. It is important to note that when the degrees of freedom are minimal, the time saving is negative. However, as the degrees of freedom increase, the time saving becomes significant, highlighting the growing efficiency of the method with the nearest neighbor search.

\section{Case-study: Self-contact in the GI system}
\label{sec:numerical application}
\subsection{Geometry and weak formulation}
\label{subsec:problem}
\begin{figure}[h!]
    \centering
    \includegraphics[width=\textwidth]{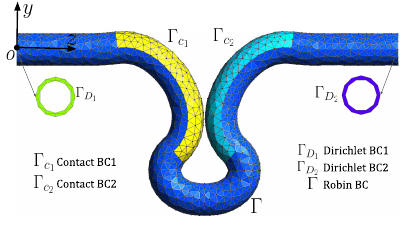}
    \caption{Problem setting: $\Gamma_{c_1}$ and $\Gamma_{c_2}$ are contact surfaces, while $\Gamma$, $\Gamma_{D_1}$, and $\Gamma_{D_2}$ are used for Robin and Dirichlet boundary conditions, respectively.}
    \label{contactconf}
\end{figure}

The model problem concerns a tract of the intestine geometrically represented as a U-shape three-dimensional domain, as shown in Fig~\ref{contactconf}. This configuration has been chosen to represent a section of the intestine that may come into self-contact during the propagation of the peristaltic wave or due to some pathology causing a dislocation from its physiological configuration.

As shown in Fig.~\ref{contactconf}, the contact surfaces are defined as the inner surfaces of the shape, represented by the colors yellow and cyan, respectively ($\Gamma_{c_1}$,$\Gamma_{c_2}$). The surfaces at the ends, colored in green and violet, will be subjected to Dirichlet-type conditions ($\Gamma_{D_1}, \Gamma_{D_2}$), assumed to be fixed or with an imposed displacement depending on the problem to be solved. The remaining boundary in blue ($\Gamma$) will be subject to Robin boundary conditions to account for the pressure exerted on the intestine by other organs, as we will discuss in the next Section.

Accordingly, the weak formulation consists in finding the displacement field $\bu$ and the hydrostatic solid pressure $p$ such that
\begin{equation}\label{contactvariational}
\begin{aligned}
\mathcal{M}(\bu,p;\delta \bu,\delta p) &:= \int_{\Omega_{0}}\bP:\nabla\delta \bu  -\int_{\Gamma_{N}} p_{0}(t)J\bF^{-T} \bn \cdot \delta \bu \\
    &+\int_{\Omega_{0}}(J-1)\delta p 
    + \int_{\Omega_{0}}\zeta_{stab}\nabla p \cdot \nabla \delta p\\
    &+\int_{\Gamma_{c_1}} K_0\langle g_{c_1}\rangle \delta \bu + 
    \int_{\Gamma_{c_2}} K_0\langle g_{c_2}\rangle \delta \bu = 0,
\end{aligned}    
\end{equation}
for all test functions $\delta \bu$ and $\delta p$. The last two integrals in \eqref{contactvariational} account for the contact forces defined in Eq. \eqref{penalty}, which are exerted between the contact surfaces $\Gamma_{c_1}$ and $\Gamma_{c_2}$ and are a result of the divergence theorem. 

\subsection{Distributed boundary stiffness}
\begin{figure}[h]
    \centering
    \includegraphics[width=1\textwidth]{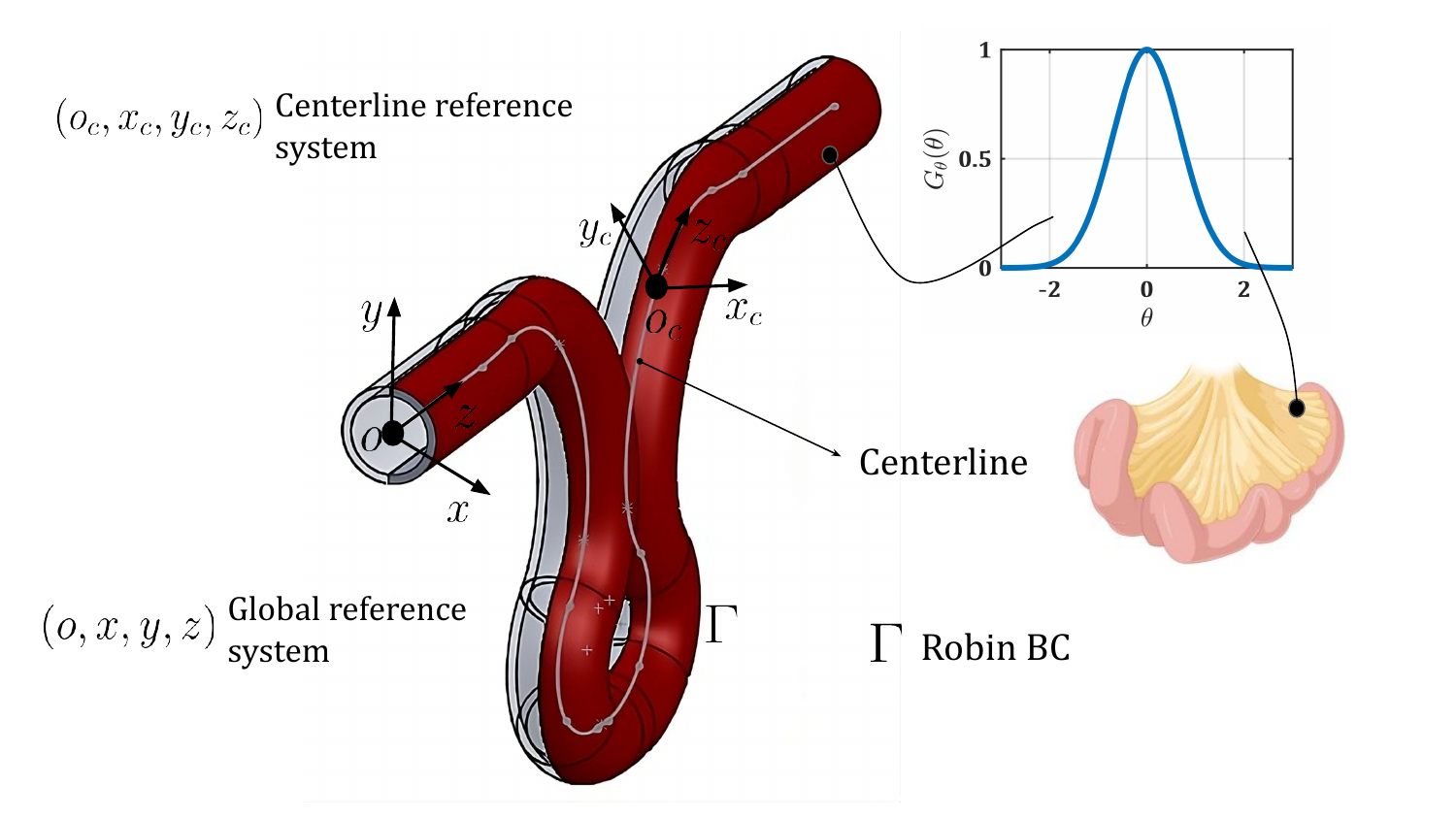}
    \caption{Schematic of the distribution of the stiffness $\eta$ on the boundary $\Gamma$. The stiffness is distributed along the surface by means of the Gaussian function $G_{\theta}(\theta)$ linked to the local reference system, $(0_c, x_c, y_c, z_c)$ on the centerline, whereas $(0, x, y, z)$ stands for the global coordinate system.}
    \label{Robin_bc}
\end{figure}

To account for distributed springs connecting the intestine with the surrounding soft tissues, we impose the following Robin boundary condition (see Fig.~\ref{Robin_bc}):
\begin{equation}
\bP + \eta(r, z) \bF^{-T} \bu = {\bf 0}, \quad \text{on } \partial \Omega \times (0, t_{\text{final}})
\label{robequ}
\end{equation}
where, the term \(\eta(r, z)\) represents a spatially varying stiffness defined as:
\begin{equation}
\label{eta}
\eta(r, s) = \eta_{min} + (\eta_{max} - \eta_{min}) \left(1 + \beta \frac{|s - s_{min}|}{|s_{max} - s_{min}|} \right) \exp\left(-\gamma \frac{r - R}{R}\right) G_{\theta}(\theta),
\end{equation}
with, $r = \sqrt{(x - x_c(s))^2 + (y - y_c(s))^2}$ the radial distance from the centerline coordinates $(x_c(s), y_c(s))$ and $s$ the curvilinear coordinate; $\eta_{min}$ and $\eta_{max}$ are the minimum and maximum stiffness values, respectively; $\beta$ introduces a stiffness gradient along the $z_c$-direction; $R$ denotes the mean radius of influence of the surrounding organs on the considered GI tract; $\gamma$ controls the radial decay of the stiffness coefficient $\eta(r, s)$ around $R$, such that the stiffness decreases progressively with distance from the mean radius. Finally, $G_{\theta}(\theta)$ is an azimuthal Gaussian function defined as:
\begin{equation}
     \label{gaussian}
     G_\theta(\theta) =
    \begin{cases}
    \exp\left(-\dfrac{(\theta - \theta_0)^2}{2\sigma^2}\right), & \text{if } |\theta - \theta_0| \leq \frac{\pi}{2}, \\
    0, & \text{else},
    \end{cases}
\end{equation}
where, the azimuthal angle $\theta$ is defined as  $\theta = \arctan2(y - y_c(s), x - x_c(s))$;
$\sigma$ stands for the standard deviation and $\theta_0$ is the preferred azimuthal angle.

The spatial distribution of $\eta(r, s)$ considers axial and radial heterogeneities in stiffness, avoiding out-of-plane displacements (see Fig.~\ref{etadis}(a)). Such an approach is inspired by \citep{propp2020orthotropic} and allows us to model several scenarios which will be analyzed in the subsequent sections:
(i) the mechanical interaction between the digestive tract and its environment (mesentery);
(ii) the presence of an herniation in a pre-surgical setting; 
(iii) the adhesion syndrome as a result of a post-surgical stiffening and the development of scar bands.

\subsection{Modeling of the presence of the mesentery}
\label{mesentery_mod}
The present case is considered as a reference for a healthy condition. The boundary parameters used to simulate the effect of the mesenteric layer attaching the intestine tract to the posterior abdominal wall are given in Tab.~\ref{table:tm}. The numerical analysis was performed maintaining the same electrophysiological and mechanical properties as those used in the case shown in Appendix~\ref{testcontactE}, where no confinement was considered. 

\begin{table}[h!]
\centering
\caption{Parameters for distributed boundaries conditions.}
\begin{tabular}{c c c c c c} 
 \hline
 $\eta_{max} \,[\rm kPa/cm ]$ & $\eta_{min} \,[\rm kPa/cm]$ & $\sigma \,[\rm rad]$ & $\beta \,[-]$ & $\gamma \,[-]$ & $\theta_0 \,[\rm rad]$ \\
 \hline
 $0.3$ & $0.1$ & $\pi/3$ & $0.5$ & $0.2$ & $0.0$\\
 \hline
\end{tabular}
\label{table:tm}
\end{table}

\begin{figure}[h]
    \centering
    \includegraphics[width=\textwidth]{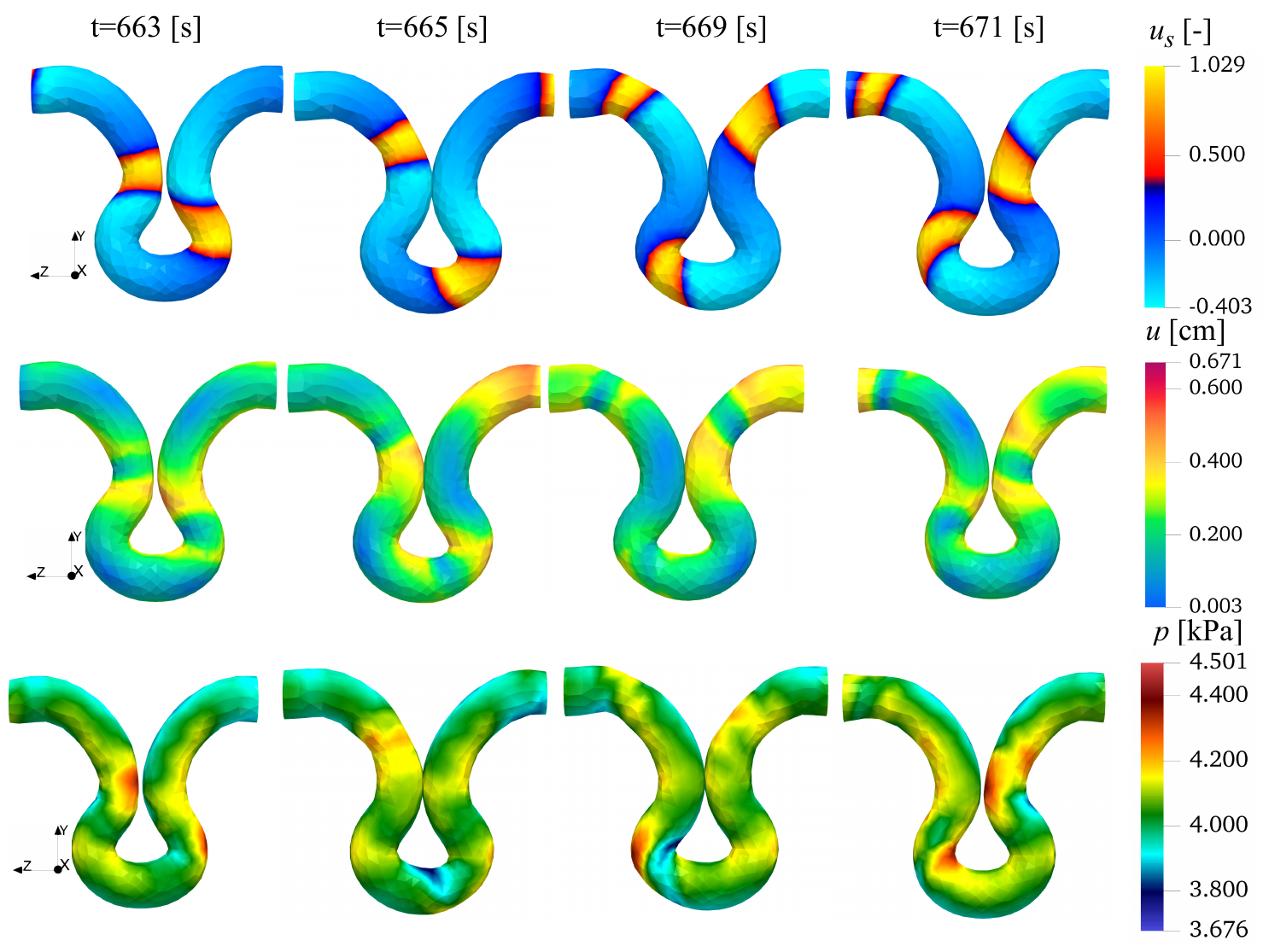}
    \caption{Temporal evolution of SMC transmembrane voltage $u_s$, hydrostatic pressure $p$ and displacement $u$. Electrophysiological parameter can be fine in \ref{sec:A}.}
    \label{ContactRobin}
\end{figure}

When comparing the two cases, due to the lack of any mechano-electric feedback, we observe the same spatiotemporal distribution of membrane potential $u_s$ (see Fig.~\ref{ContactRobin}). 

However, a notable difference in the global displacement $\bu$ is obtained. In the unconfined case, Fig.~\ref{testcontactE}, the displacement reaches a maximum of  ~1.517 cm. In contrast, with the application of the Robin-type condition, the global displacement is reduced by approximately ~0.671 cm (see Fig.~\ref{ContactRobin}). This demonstrates that the condition effectively maintains a certain level of intestinal stability.

\begin{figure}[h!]
    \centering
    \includegraphics[width=\textwidth]{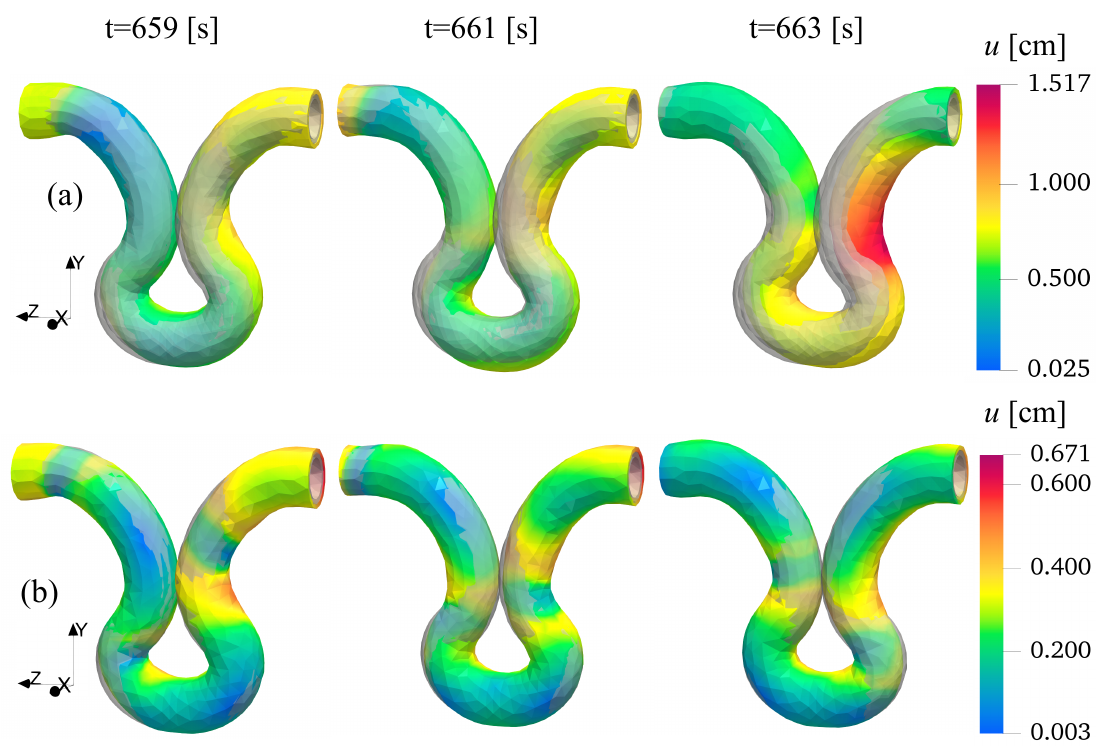}
    \caption{Effect of the Robin boundary condition. Displacement magnitude $\bu$ (a) without and (b) with the Robin boundary condition applied.}
    \label{Comparaison}
\end{figure}

This conclusion is further supported by the pressure distribution curves, which show higher pressure in the unconfined case (Fig.~\ref{testcontactE}). Indeed, without a distributed boundary condition, the intestine moves more freely, leading to increased pressure. To better illustrate our findings, Fig.~\ref{Comparaison} compares the displacement configurations among the two cases at t=659s, t=661s, and t=663s. The results clearly show that when the intestine is not confined by the presence of other organs, it exhibits significant motility, even in the out of the plane direction, as highlighted in Fig.~\ref{Comparaison}(a). In contrast, when the Robin boundary condition is applied, the domain is stabilized and prevented from moving freely in all directions, as shown in Fig.~\ref{Comparaison}(b).

\begin{figure}[h]
    \centering
    \includegraphics[width=0.8\textwidth]{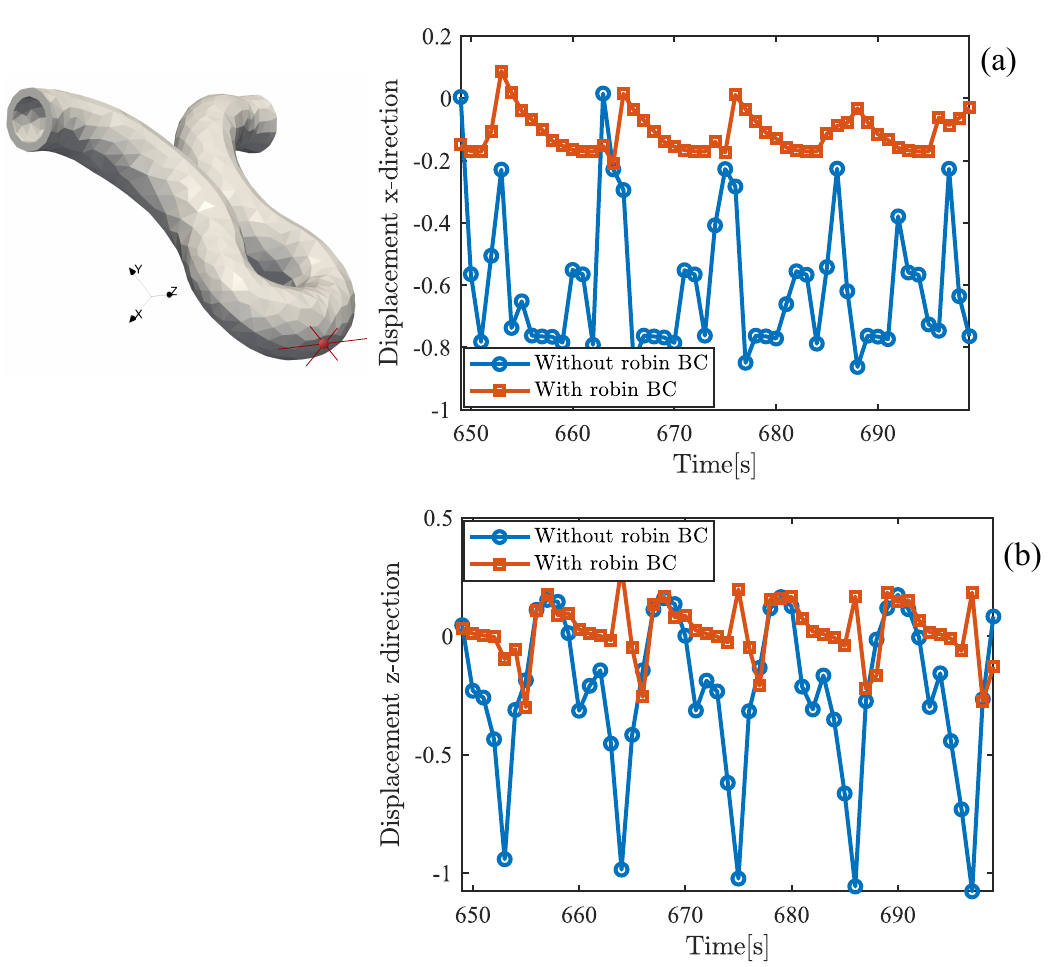}
    \caption{Comparison of displacement components time course at point $(0, -17, 19.4)$:
    (a) $x-$direction, and (b) $z-$direction.}
    \label{qualitative}
\end{figure}

Figure~\ref{qualitative} confirms quantitatively such an observation, showing the displacements in the $x$- (Fig.~\ref{qualitative}(a))  and $z$- (Fig.~\ref{qualitative}(b)) directions for the two cases, measured at a point with coordinates $(0, -17, 19.4)$ (represented in red in Fig.~\ref{qualitative}). It is clear that when the Robin boundary condition is absent, a larger displacement with associated fluctuations is obtained.

\begin{figure}[h]
    \centering
    \includegraphics[width=0.65\textwidth]{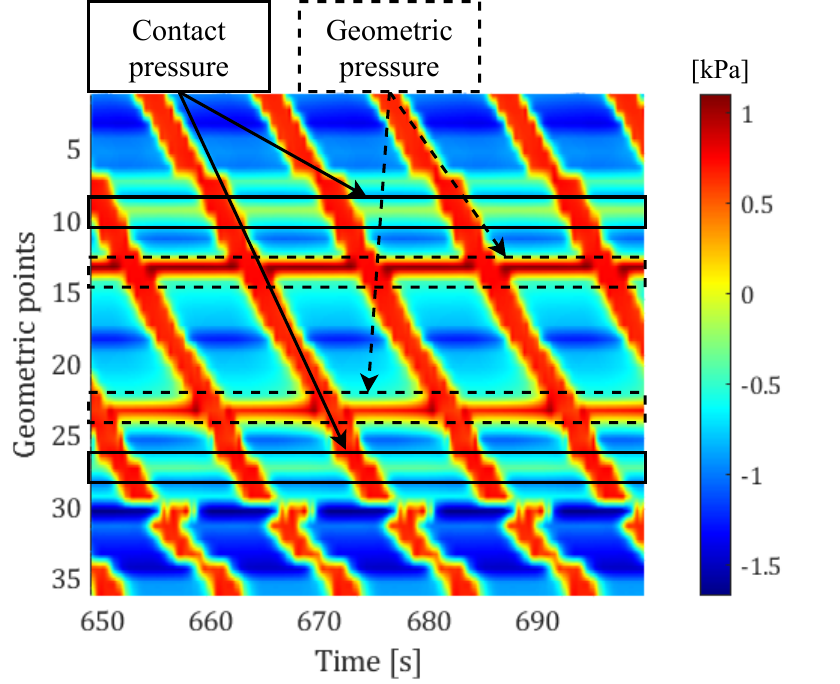}
    \caption{Simulated manometry for the healthy case. The regions labeled as `Geometric pressure' represent constant pressures induced by the corners of the geometry while the regions labeled `Contact pressure' are due to the pressure exerted by self-contact.}
    \label{manohealty}
\end{figure}

Figure~\ref{manohealty} shows the intraluminal pressure map evaluated on geometrical points along the tract (manometry patterns) for the case with the Robin boundary conditions, serving as a baseline for comparing subsequent results. Regions labeled as `Geometric pressure' represent constant pressure values induced by the corners of the geometry. Regions labeled `Contact pressure' are due to the pressure exerted by the self-contact boundary. While these pressures are essential for understanding the structure of the problem, they do not significantly affect the peristaltic motion, as confirmed by the diagonal patterns in Fig.~\ref{manohealty}. However, it is worth noticing that the model can highlight areas characterized by pressure concentrations and the pressures induced by contact between the intestinal surfaces.
These results demonstrate that our Robin-type boundary condition successfully replicates the stabilizing role of the mesentery and surrounding organs.

\clearpage

\subsection{Modeling strangulation in abdominal hernia}
\begin{figure}[h]
    \centering
    \includegraphics[width=0.5\textwidth]{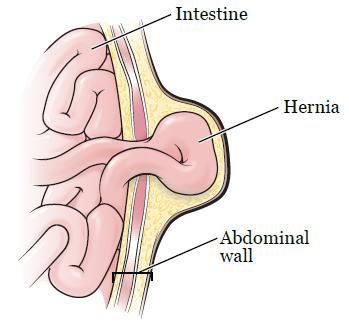}
    \caption{An abdominal strangulated hernia (picture adapted from \citep{mskcc_hernia_surgery}).}
    \label{Hernia}
\end{figure}

We consider intestinal hernias as a clinical case study (see Fig.~\ref{Hernia}). Intestinal hernia is a pathological condition in which part of the intestine protrudes through an opening in the abdominal wall or a weakened muscle. Such a condition can occur in a number of anatomical locations, notably in the groin (inguinal hernia) or around the navel (umbilical hernia). Two levels of herniation are considered, moderate and severe \citep{wiesner2011small, slater2014criteria, csimcsek2020factors}.

\begin{figure}[h]
    \centering
    \includegraphics[width=0.95\textwidth]{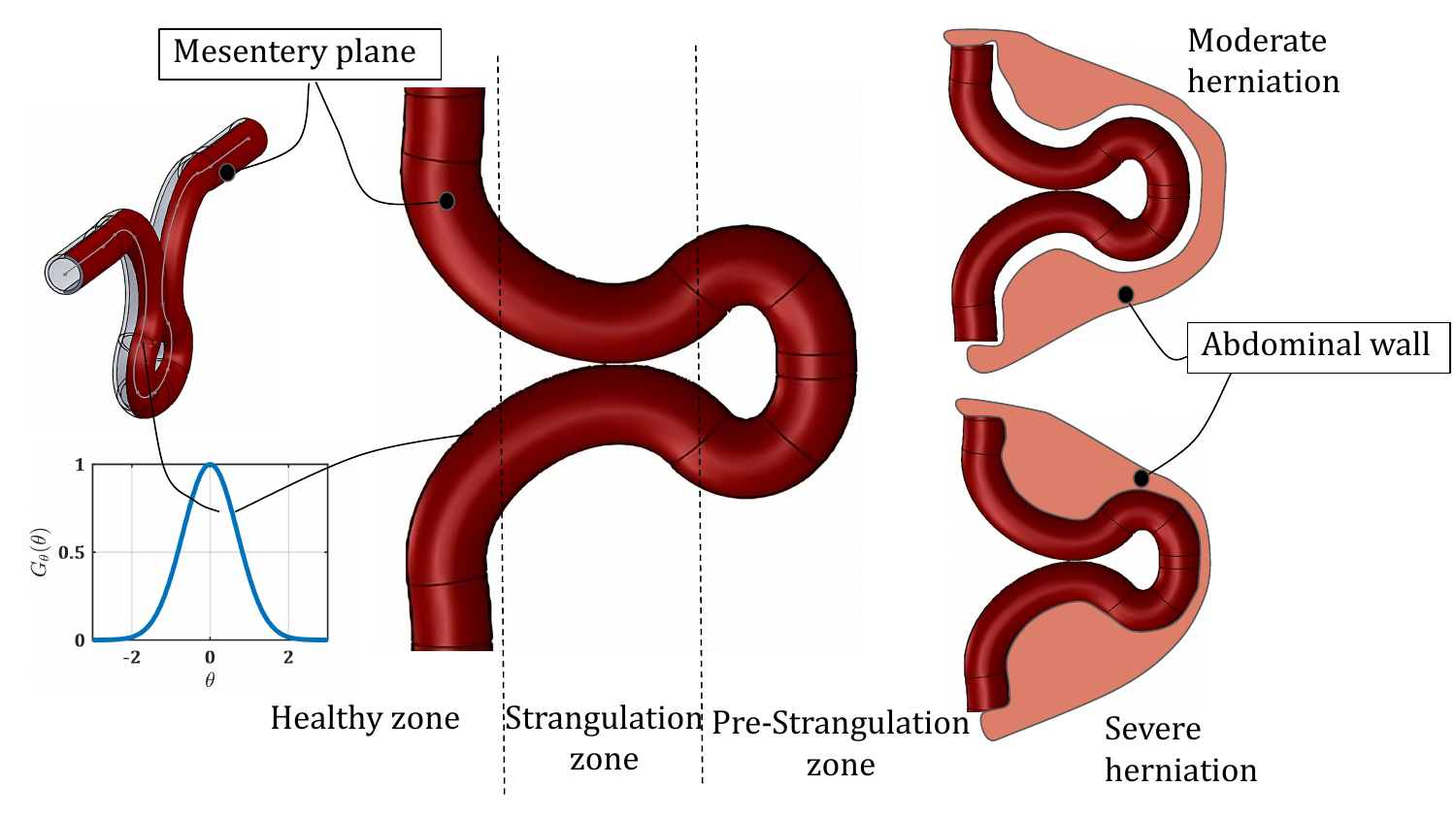}
    \caption{Hernia problem configuration schematics. The mesentery boundary is highlighted in red with associated Gaussian function $G_\theta(\theta)$. Healthy section, strangulation and pre-strangulation zones are identified. Moderate and severe herniation cases are provided on the right.}
    \label{hernia_conf}
\end{figure}

The problem configuration is similar to that explained in Subsection~\ref{mesentery_mod}. The area affected by a hernia can be divided into three distinct zones (as shown in Fig. \ref{hernia_conf}), each with specific mechanical properties. In the healthy zone, the intestinal wall retains its physiological elastic characteristics. In the strangulation zone, blood flow is interrupted, leading to ischemia, degradation of cellular structures, and a significant increase in stiffness. Finally, the pre-strangulation zone is subjected to mechanical stress with possible signs of inflammation and edema, leading to an intermediate stiffness between the healthy zone and the strangulation zone, influenced by fluid accumulation and partial tissue degradation \citep{wiesner2011small, sghaier2023extensive, csimcsek2020factors}. Simulation parameters are given in Tab.~\ref{tab:herniaP}.

In the case of severe intestinal hernia, local variations in electrophysiological properties are observed \citep{misiakos2014strangulated, keeley2019predictors, csimcsek2020factors}. The healthy zone, away from the compression, retains physiological behavior with intact conductivity. The pre-strangulation zone undergoes mechanical stress and partial impairment of blood perfusion, resulting in mild inflammation and edema. These effects locally increase the electrical resistance, modeled by a moderate reduction in the diffusion coefficient. Finally, the strangulation zone is marked by severe ischemia and cellular necrosis. This region is modeled by a drastic decrease in the diffusion coefficient, representing a significant reduction in the conduction capacity of electrical signals. The list of modified parameters is provided in \ref{sec:A}.

\begin{table}[h!]
\centering
\caption{Mechanical constitutive parameters for Hernia simulation.}
\begin{tabular}{|c|c|c|c|}
\hline
\textbf{Zones} & \textbf{Healthy} & \textbf{Strangulation} & \textbf{Pre-Strangulation} \\ \hline
$\mu \,[\rm kPa ]$ & $2.5$ & $3$  & $2.7$ \\ \hline
$k_1^l \,[\rm kPa]$  & $5.14$ & $7$ & $6$ \\ \hline
$k_2^l \,[-]$ & $1.19$ & $2.1$ & $2$ \\ \hline
$k_1^c \,[\rm kPa]$& $0.78$ & $0.9$ & $0.8$ \\ \hline
$k_2^c \,[-]$ & $0.02$ & $0.04$ & $0.03$ \\ \hline
$k_1^d \,[\rm kPa]$ & $3.65$ & $4$ & $3.66$ \\ \hline
$k_2^c \,[-]$  & $0.31$ & $0.35$ & $0.33$ \\ \hline
\end{tabular}
\label{tab:herniaP}
\end{table}

\begin{figure}[h!]
    \centering
    \includegraphics[width=\textwidth]{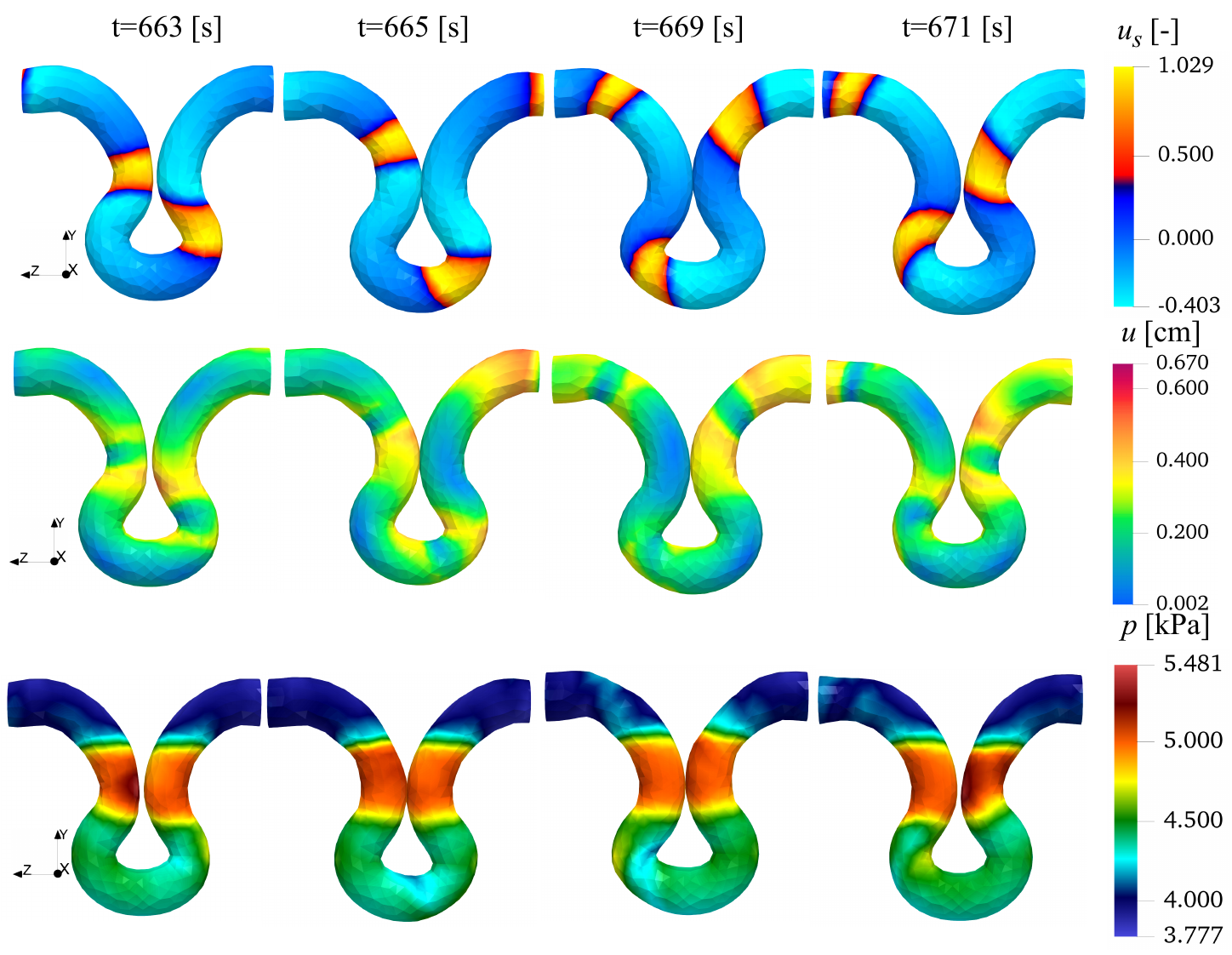}
    \caption{Temporal evolution of SMC transmembrane voltage $u_s$, hydrostatic pressure $p$ and displacement $u$ for the moderate hernia case. Electrophysiological parameters are in \ref{sec:A}.}
    \label{Contacthernia}
\end{figure}

\begin{figure}[h!]
    \centering
    \includegraphics[width=\textwidth]{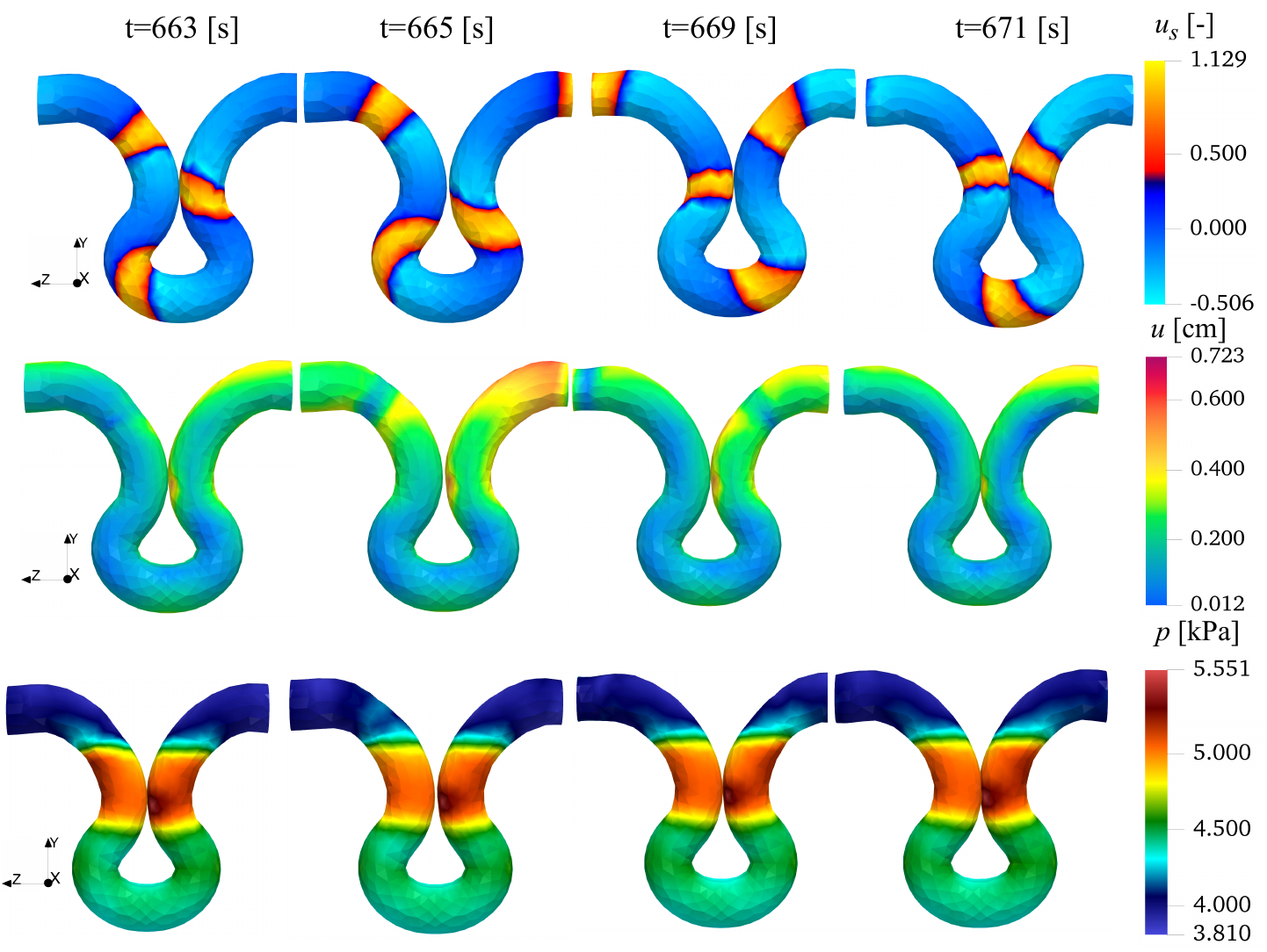}
    \caption{Temporal evolution of SMC transmembrane voltage $u_s$, hydrostatic pressure $p$ and displacement $u$ for the severe hernia case. Electrophysiological parameters can be find in \ref{sec:A}.}
    \label{ContactherniaDiff}
\end{figure}

\begin{figure}[h!]
    \centering
    \includegraphics[width=1.01\textwidth]{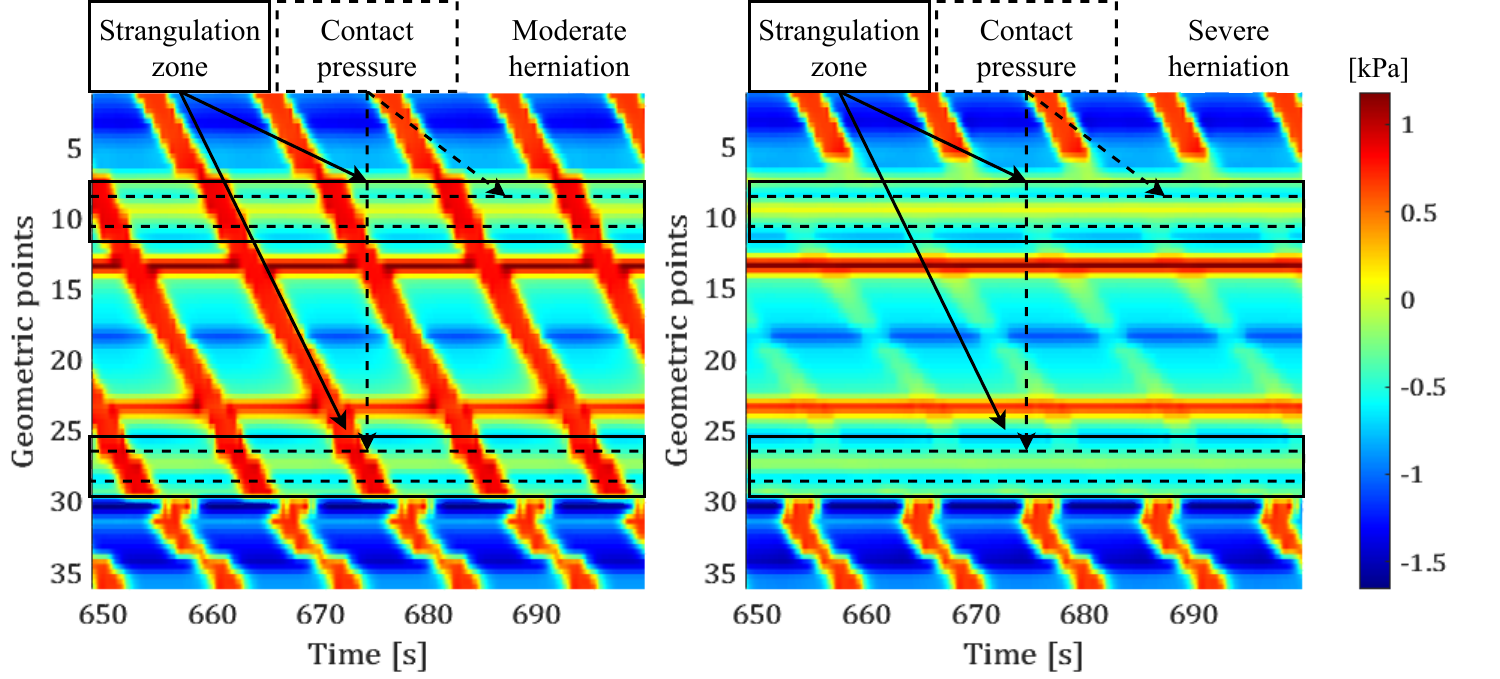}
    \caption{Simulated manometry in the moderate (left) and severe (right) hernia region. In the strangulation zone, no contractions are observed for the severe case.}
    \label{Cmanoherniadiff}
\end{figure}

Simulation results for moderate strangulation are provided in Fig.~\ref{Contacthernia}: the electrical wave propagates without significant constraints, allowing almost physiological intestine contraction in the strangulation and pre-strangulation zones. The overall motility of the healthy zone is maintained, thus remaining functional. However, in the case of severe strangulation, Fig.~\ref{ContactherniaDiff}, a marked slowing of the electrical wave is observed in the strangulation and pre-strangulation zones, leading to a significant reduction in contraction rate and associated motility. 

Manometric measurements in strangulation hernias are technically impossible to perform in clinical conditions due to the severe compression of the intestinal lumen that prevents catheter passage. In our self-contact computational framework, intraluminal pressure levels can be analyzed for the first time. In the case of moderate strangulation, Fig.~\ref{Cmanoherniadiff}, the manometric curves showed active contraction in the pre-strangulation zone, constant pressure in the strangulation zone (attributed to pressure exerted by contact surfaces), and persistent physiological contraction in the healthy zones. On the other hand, in the case of severe strangulation, Fig.~\ref{Cmanoherniadiff}, the curves also revealed a constant contact pressure in the strangulation zone and absence of contraction due to the contact surfaces, associated with negligible contraction in the pre-strangulation zone. It is worth noticing that, in the severe case, the constant pressure is more pronounced than in the moderate case. Such high stress is due to the additional load exerted by the abdomen, acting as an external constraint.

\begin{figure}[h]
    \centering
    \includegraphics[width=\textwidth]{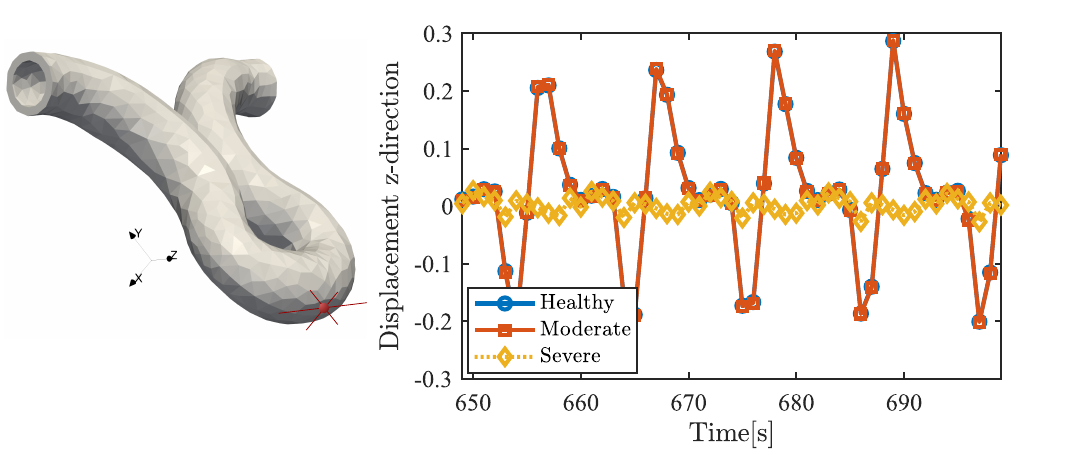}
    \caption{Comparison of $z-$direction displacement time course between health, moderate, and severe hernia configurations at point $(0, -17, 19.4)$.}
    \label{qualitativeH}
\end{figure}

To better understand the underlying phenomena, Fig.~\ref{qualitativeH} illustrates the displacement in the $z$-direction within the pre-strangulation zone for three scenarios: healthy, moderate hernia, and severe hernia. For the moderate hernia case, the displacement closely resembles the healthy case, suggesting the onset of a hernia or a very mild condition. On the other hand, in the case of severe hernia, the computed displacement is significantly smaller despite muscle contraction responding to electrical activation. Such a lack of motility can be attributed to the portion of the intestine that has passed through and is constrained by the abdomen. 
\newpage

\subsection{Modeling intestinal adhesion syndrome}
One potential application of our model is the study of intestinal adhesion syndrome \citep{tabibian2017abdominal}, a common condition that arises after surgical interventions or abdominal inflammations. This syndrome is characterized by the formation of fibrous bands, known as adherents, which connect different parts of the intestine or adjacent organs. These adhesions can lead to severe complications such as chronic pain, intestinal obstructions, and impaired intestinal motility \citep{attard2007adhesive, strik2016long}.

\begin{figure}[h]
    \centering
    \includegraphics[width=0.95\textwidth]{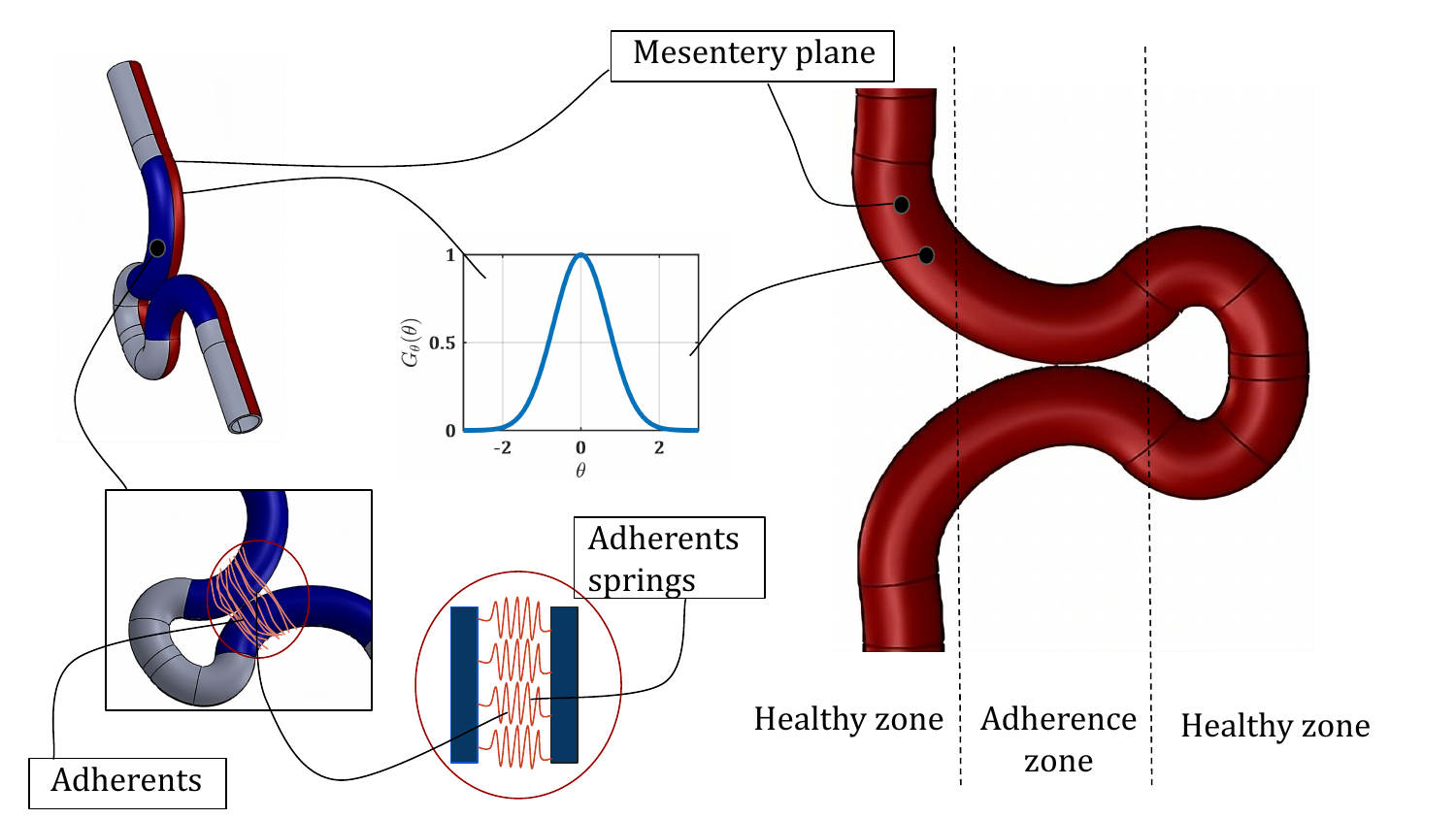}
    \caption{Adhesion problem configuration schematics. The mesentery surface is highlighted in red with the associated Gaussian distribution of stiffness. Adherents are represented as distributed springs on the blue boundary.}
    \label{Adherence}
\end{figure}

To investigate this phenomenon without directly modeling the adhesions themselves, we based our study on the geometry shown in Fig.~\ref{Adherence}. From this configuration, we aimed to simulate the restrictive effects of adhesions by introducing specialized boundary conditions as show in Eq.~\ref{robequA}.
\begin{equation}
\bP + \eta(r, z) \bF^{-T} \bu + \eta_a \bF^{-T} (\bu-\bu_{ref})= 0, \quad \text{on } \partial \Omega \times (0, t_{\text{final}})
\label{robequA}
\end{equation}
here, $\eta_a$ stands for the adherence stiffness (see Fig.~\ref{etadis}(b)) while $\bu_{ref}$ is the initial displacement of the intestinal wall. This approach involves adding a mechanical constraint that replicates the impact of adhesions on displacements and forces. Through this method, we can examine how localazied restrictions influence intestinal dynamics while maintaining a general electromechanical framework. Material stiffness changes according to Tab.~\ref{tab:herniaP}.

\begin{figure}[h!]
    \centering
    \includegraphics[width=0.65\textwidth]{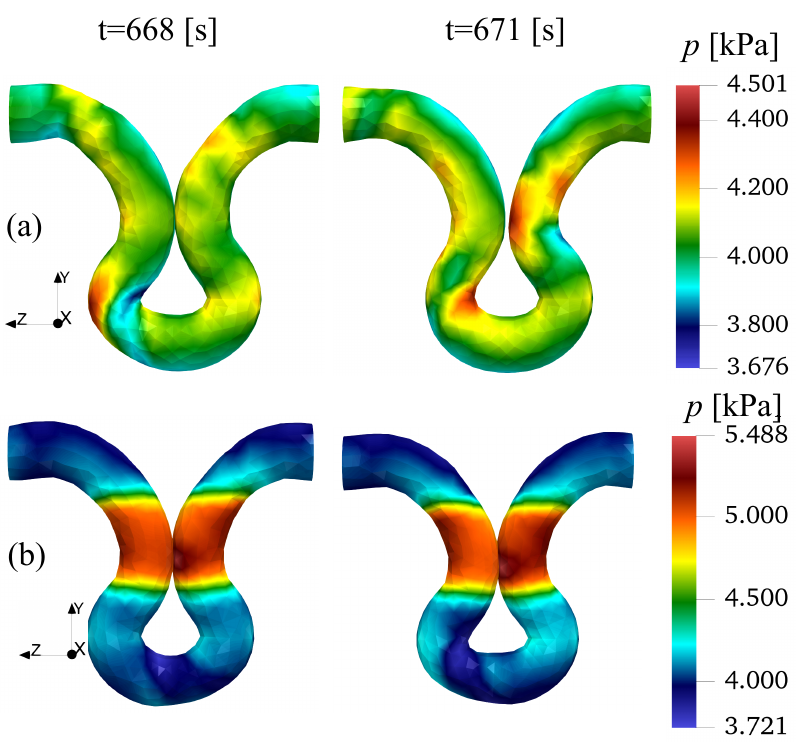}
    \caption{Temporal evolution of hydrostatic pressure $p$ for the severe intestinal adhesion syndrome: (a) case without adhesion, and (b) case with adhesion, where material stiffness has been modified in the adhesion region: $\eta_a = 0.9 \,[\rm kPa/cm]$, $\bu_{ref} = 0$; electrophysiological parameters can be find in \ref{sec:A}.}
    \label{ContactAdh}
\end{figure}

\begin{figure}[h!]
    \centering
    \includegraphics[width=0.65\textwidth]{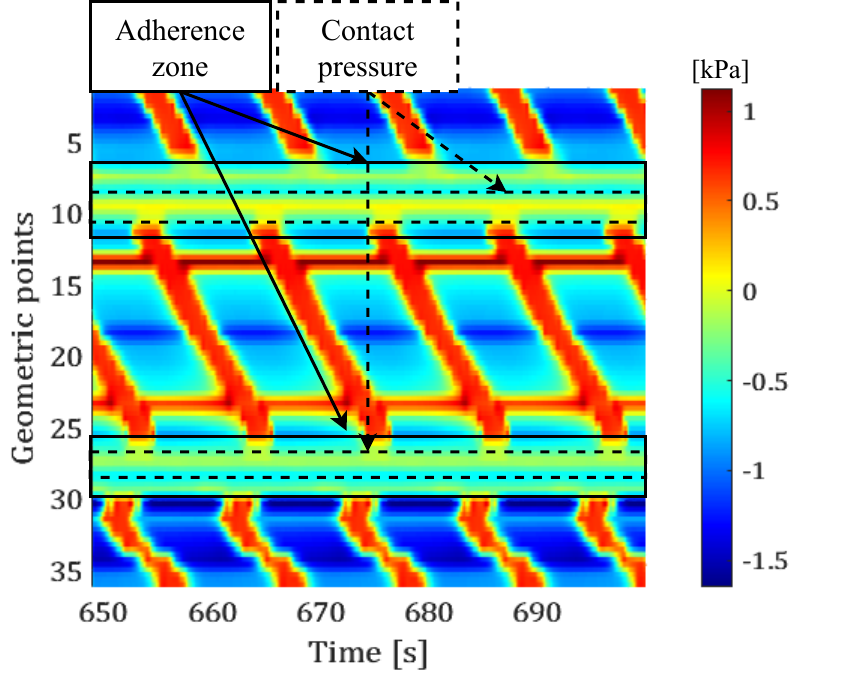}
    \caption{Simulated manometry in the adhesion syndrome region. In the adherence zone, no contractions are observed. The area referred to as `Contact pressure' shows the pressure levels due to self-contact surfaces.}
    \label{manoADH}
\end{figure}

\begin{figure}[h!]
    \centering
    \includegraphics[width=\textwidth]{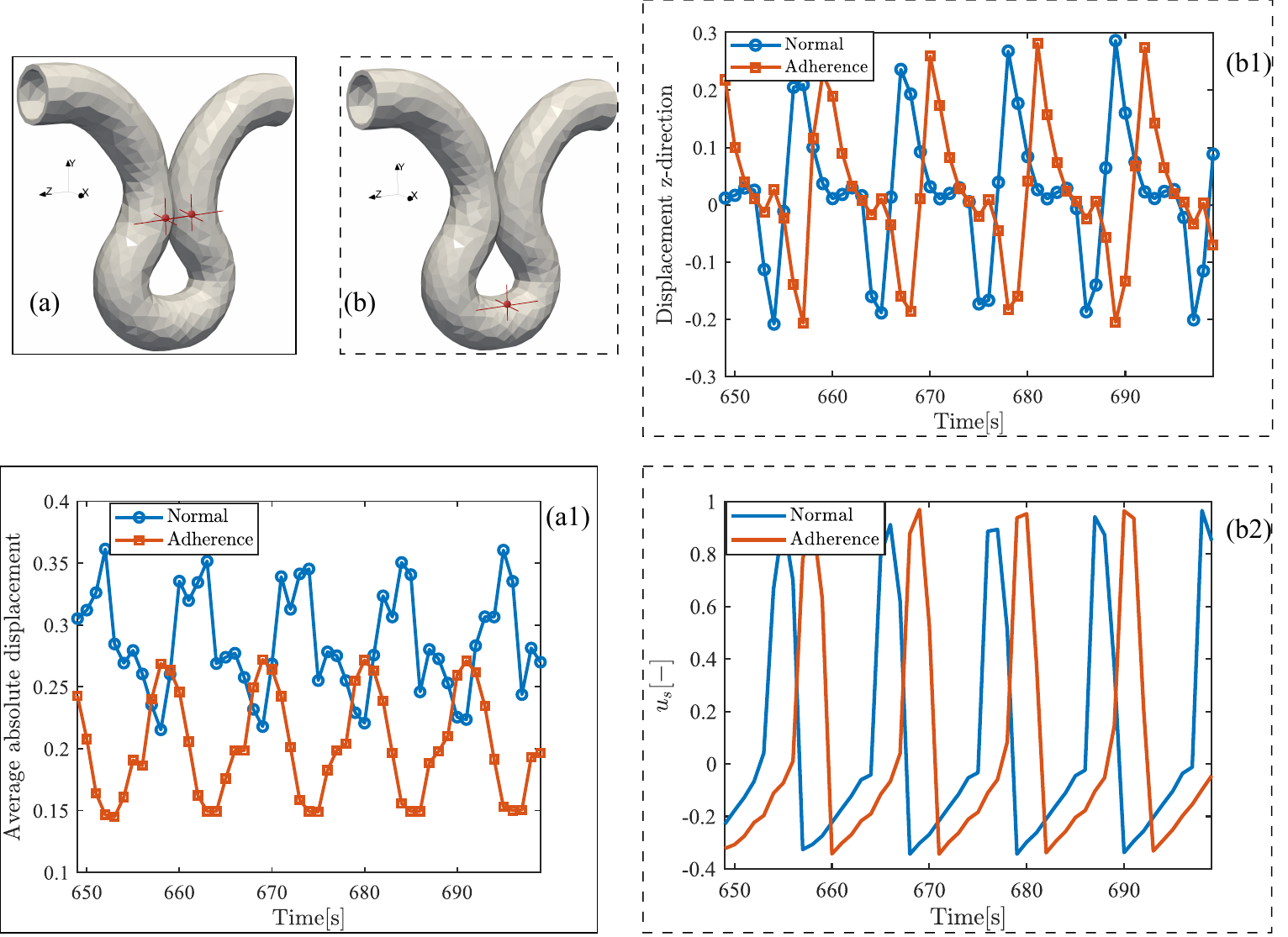}
    \caption{(a) Absolute displacement between two points taking on the adherence region; (b) Comparison of the displacement and smooth muscle membrane potential of a point coordinate $(1.40, -14.1, 18.9)$ after adhesion region: (b1) membrane potential $u_s$ in $x-$direction, and (b2) displacement $u$ in $z-$direction. 
    }
    \label{ADcompar}
\end{figure}

\begin{table}[h!]
\centering
\caption{Mechanical constitutive parameters for Hernia simulation}
\begin{tabular}{|l|c|c|}
\hline
\textbf{Zones} & \textbf{Healthy} & \textbf{Adherence}  \\ \hline
$\mu \,[\rm kPa ]$ & $2.5$ & $3.1$  \\ \hline
$k_1^l \,[\rm kPa]$  & $5.14$ & $7$ \\ \hline
$k_2^l \,[-]$ & $1.19$ & $2.1$  \\ \hline
$k_1^c \,[\rm kPa]$& $0.78$ & $0.8$  \\ \hline
$k_2^c \,[-]$ & $0.02$ & $0.04$ \\ \hline
$k_1^d \,[\rm kPa]$ & $3.65$ & $4$  \\ \hline
$k_2^c \,[-]$  & $0.31$ & $0.35$  \\ \hline
\end{tabular}
\label{tab:herniaP}
\end{table}
For this last test case, we present only the pressure maps in Fig.~\ref{ContactAdh}, comparing them with the results from Fig.~\ref{ContactRobin}, where adhesion was not considered. At $t=668$ s, in both cases, self contact is observed. However, at $t=671 $ s, a clear difference emerges: in the scenario without adhesion, the intestinal surfaces separate due to the peristaltic wave, while in the case with adhesion, the surfaces remain in contact. Such a condition is further confirmed in Fig.~\ref{ADcompar} showing the average absolute displacement between two points located at the adherence region. The absolute displacement is reduced then adhesions are present. Reduced motility is also observed on the manometry curves in Fig.~\ref{manoADH}. In particular, instead of contraction waves, a slight constant pressure appears, indicating the force exerted by the adhesion, forcing the intestinal walls to remain in contact.

To further investigate this condition, we examined the behavior beyond the adhesion zone. Fig.~\ref{ADcompar}(b1) and Figure Fig.~\ref{ADcompar}(b2) display the displacement and action potential at a point located after the adhesion zone. We observe that the action potential is delayed in the case with adhesion, which consequently slows down the contraction speed. However, this does not systematically affect the displacement amplitude. These findings are in perfect agreement with the manometry results.

\clearpage
\newpage

\section{Conclusions}
\label{sec:conclusion}

The study proposes a comprehensive multiphysics framework for modeling intestinal motility, integrating electromechanics, tissue anisotropy, cellular electrophysiology, and self-contact extending and generalizing previous works from the authors \cite{DJOUMESSI2024116989}. The contact method developed is based on distance computation between contact surfaces employing a penalty force to prevent penetration. To reduce the computational time for evaluating the gap between contact surfaces, an enhanced search method based on nearest-neighbor algorithms was implemented. The method was validated using a benchmark test, quantifying the gap violations at the contact interface. Results demonstrated minimal gap violations, showcasing the robustness of the approach. The governing equations were discretized using the finite element method, and a custom code leveraging on \texttt{FEniCS} and \texttt{Gmsh} was developed for their resolution.

We used the model to study the influence of surrounding organs on intestinal motility. For this purpose, a specialized boundary condition was developed, incorporating a progressive stiffness distribution for external organs (e.g., the mesentery) combined with Gaussian functions. Results indicated that these conditions play a critical role in constraining intestinal motion, preventing free movement. Finally, we investigated two pathological conditions: abdominal hernia and abdominal adhesion (post surgical situation). Although reproducing manometry in such pathological conditions is challenging to implement in clinical practice, our model allowed us to simulate manometric curves in these different pathological scenarios, demonstrating its potential as a predictive tool for clinical applications.

This work investigates the in silico prediction of intestinal electromechanical motility in curved three-dimensional geometries susceptible to self-contact for the first time. Demonstrated model predictability averages several applications in a clinical scenario.

\paragraph{Limitations and Perspectives}
Model limitations are briefly mentioned. The implemented electrophysiological model can reproduce one type of excitation wave at a time, either slow waves or spikes. Moreover, action potential propagation is unidirectional (adoral direction) not considering the enteric nervous system~\citep{barth2017electrical, barth2018computational,athavale2024mapping}. The mesentery was reproduced by introducing a Robin-type boundary condition with a spatial distribution of stiffness. Advanced modeling approaches will require porous interfaces enabling multiphysics flux exchanges with the surrounding organs~\citep{lourencco2022poroelastic, barnafi2022finite, careaga2024new}. From a numerical perspective, a contact formulation based on the penalty method can become unstable if the penalty parameter is not properly chosen. As future development, we plan to introduce an adaptive penalty coefficient to enhance model robustness for large scale numerical analyses.




\section*{Acknowledgement}
Authors acknowledge the support of the Italian National Group for Mathematical Physics (GNFM-INdAM).

\bibliographystyle{elsarticle-num-names}
\bibliography{sample}

\clearpage
\newpage
\appendix
\section{Parameters for the Electrophysiology for strangulation hernia and Adhesion symdrome}
\label{sec:A}
\begin{table}[h!]
\centering
\caption{Electrophysiological parameters adapted in Healthy zone from \citep{aliev2000simple,gizzi2010electrical}.}
\begin{tabular}{c c c c} 
 \hline
\multicolumn{2}{c}{SMC layer} &\multicolumn{2}{c}{ICC layer} \\
 \hline
 $k_s$=$10$ & $a_s$=$0.06$ &$k_i$=$7$ & $a_i$=$0.5$ \\ 
 $\beta_s$ = $0$ & $\lambda_s$=$8$ & $\beta_s$ = $0.5$ & $\lambda_i$=$8$ \\
 $\epsilon_s$=$0.15$ & $\alpha_s$=$1$ & $\epsilon_i$=$\epsilon_i(z)$ & $\alpha_i$=$-1$ \\
 $D_{si}$=$0.3$ & $D_s$=$0.4$ & $D_{is}$=$0.3$ & $D_i$=$0.04$ \\
 \hline
\end{tabular}
\label{table:1}
\end{table}

In the hernia, for the strangulation zone, the diffusion coefficients $D_s^s=0.1D_s$ and $D_i^s=0.1D_i$ and in the pre-strangulation, the diffusion coefficients $D_s^{ps}=0.5D_s$ and $D_i^{ps}=0.5D_i$.

For the Adhesion, the diffusion coefficients are $D_s^a=0.125D_s$ and $D_i^a=0.125D_i$ in the adhesion region.

\section{Distribution of the stiffness for Robin Bc used in all simulations}
\label{sec:B}
This section illustrates the distribution of stiffness $\eta(r, s)$ across a surface of the geometry and the additional stiffness $\eta_a$, which accounts for the effects of adhesions. Since the stiffness $\eta(r, s)$ represents the influence of the mesentery, it was included in all simulations. On top of this baseline stiffness, an additional stiffness was introduced solely in the presence of adhesions to simulate the case of the adhesion syndrome.
\begin{figure}[h!]
    \centering
    \includegraphics[width=1.1\textwidth]{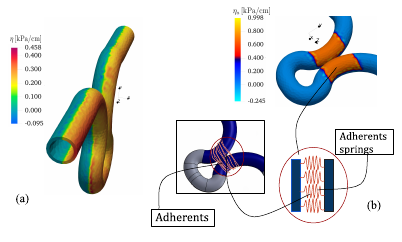}
    \caption{(a) distribution of the stiffness $\eta$ representing the effect of mesentery and (b) show the distribution of the stiffness $\eta_a$ showing the effect of the adherents.}
    \label{etadis}
\end{figure}

\newpage
\section{Application to the problem setting}
\label{sec:application}

After testing our contact code on the benchmark problem (see Fig.~\ref{benchmarkresult1}), we proceeded to integrate the contact code into the full electromechanical framework and solved the problem on the geometry described in Fig.~\ref{contactconf} using the mechanical problem explained in Eq.~\ref{contactvariational}.

\begin{figure}[h!]
    \centering
    \includegraphics[width=\textwidth]{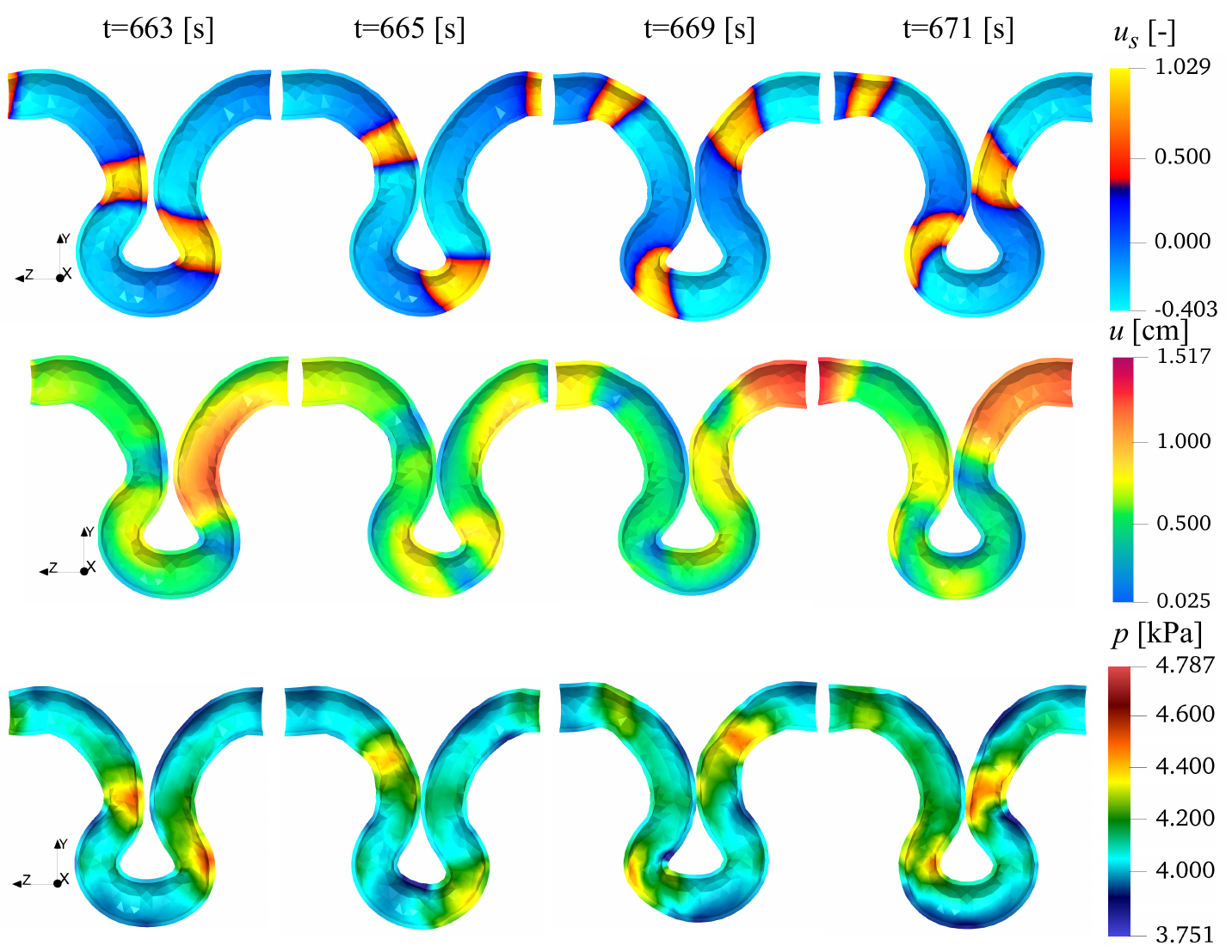}
    \caption{Temporal evolution of SMC transmembrane voltage $u_s$, of hydrostatic pressure $p$ and the displacement $u$ for the Euclidean distance method}
    \label{testcontactE}
\end{figure}

We observe in Fig.~\ref{testcontactE} that the contact code effectively prevents interpenetration between the contact surfaces. This is particularly evident in the pressure curves, which show an increase in pressure at the contact regions when the surfaces come into contact, notably at t=667s and t=669s. This confirms that the contact between the two surfaces is successfully handled by the implemented code.This result was used as a reference for the rest of the simulation

\end{document}